\documentclass[12pt]{article}
\usepackage[T1]{fontenc}
\usepackage[utf8]{inputenc}
\usepackage{geometry}
\usepackage{color}
\usepackage{mathtools}
\usepackage{amsmath}
\usepackage{amssymb}
\usepackage{graphicx}
\usepackage{setspace}
\usepackage[authoryear]{natbib}
\usepackage{breakurl}

\makeatletter
\usepackage{chngcntr}
\usepackage{mathtools}
\counterwithout{figure}{section}
\counterwithout{equation}{section}
\usepackage{times}
\usepackage{subfig}
\usepackage{floatrow}

\date{}

\makeatother

\begin{document}
\global\long\def\jm{J_{m}}
 \global\long\def\dg{\mathbf{F}}
 \global\long\def\dgcomp#1{F_{#1}}
 \global\long\def\piola{\mathbf{P}}
 \global\long\def\refbody{\Omega_{0}}
 \global\long\def\refbnd{\partial\refbody}
 \global\long\def\bnd{\partial\Omega}
 \global\long\def\rcg{\mathbf{C}}
 \global\long\def\lcg{\mathbf{b}}
 \global\long\def\rcgcomp#1{C_{#1}}
 \global\long\def\cronck#1{\delta_{#1}}
 \global\long\def\lcgcomp#1{b_{#1}}
 \global\long\def\deformation{\boldsymbol{\chi}}
 \global\long\def\dgt{\dg^{\mathrm{T}}}
 \global\long\def\idgcomp#1{F_{#1}^{-1}}
 \global\long\def\velocity{\mathbf{v}}
 \global\long\def\accel{\mathbf{a}}
 \global\long\def\vg{\mathbf{l}}
 \global\long\def\idg{\dg^{-1}}
 \global\long\def\cauchycomp#1{\sigma_{#1}}
 \global\long\def\idgt{\dg^{\mathrm{-T}}}
 \global\long\def\cauchy{\boldsymbol{\sigma}}
 \global\long\def\normal{\mathbf{n}}
 \global\long\def\normall{\mathbf{N}}
 \global\long\def\traction{\mathbf{t}}
 \global\long\def\tractionl{\mathbf{t}_{L}}
 \global\long\def\ed{\mathbf{d}}
 \global\long\def\edcomp#1{d_{#1}}
 \global\long\def\edl{\mathbf{D}}
 \global\long\def\edlcomp#1{D_{#1}}
 \global\long\def\ef{\mathbf{e}}
 \global\long\def\efcomp#1{e_{#1}}
 \global\long\def\efl{\mathbf{E}}
 \global\long\def\freech{q_{e}}
 \global\long\def\surfacech{w_{e}}
 \global\long\def\outer#1{#1^{\star}}
 \global\long\def\perm{\epsilon_{0}}
 \global\long\def\matper{\epsilon}
 \global\long\def\jump#1{\llbracket#1\rrbracket}
 \global\long\def\identity{\mathbf{I}}
 \global\long\def\area{\mathrm{d}a}
 \global\long\def\areal{\mathrm{d}A}
 \global\long\def\refsys{\mathbf{X}}
 \global\long\def\Grad{\nabla_{\refsys}}
 \global\long\def\grad{\nabla}
 \global\long\def\divg{\nabla\cdot}
 \global\long\def\Div{\nabla_{\refsys}}
 \global\long\def\derivative#1#2{\frac{\partial#1}{\partial#2}}
 \global\long\def\aef{\Psi}
 \global\long\def\dltendl{\edl\otimes\edl}
 \global\long\def\tr#1{\mathrm{tr}#1}
 \global\long\def\ii#1{I_{#1}}
 \global\long\def\dh{\hat{D}}
 \global\long\def\inc#1{\dot{#1}}
 \global\long\def\sys{\mathbf{x}}
 \global\long\def\curl{\nabla}
 \global\long\def\Curl{\nabla_{\refsys}}
 \global\long\def\piolaincpush{\boldsymbol{\Sigma}}
 \global\long\def\piolaincpushcomp#1{\Sigma_{#1}}
 \global\long\def\edlincpush{\check{\mathbf{d}}}
 \global\long\def\edlincpushcomp#1{\check{d}_{#1}}
 \global\long\def\efincpush{\check{\mathbf{e}}}
 \global\long\def\efincpushcomp#1{\check{e}_{#1}}
 \global\long\def\elaspush{\boldsymbol{\mathcal{C}}}
 \global\long\def\elecpush{\boldsymbol{\mathcal{A}}}
 \global\long\def\elaselecpush{\boldsymbol{\mathcal{B}}}
 \global\long\def\disgrad{\mathbf{h}}
 \global\long\def\disgradcomp#1{h_{#1}}
 \global\long\def\trans#1{#1^{\mathrm{T}}}
 \global\long\def\q#1{#1^{\left(q\right)}}
\global\long\def\phase#1{#1^{\left(n\right)}}
\global\long\def\n#1{#1^{\left(n\right)}}
\global\long\def\m#1{#1^{\left(m\right)}}
\global\long\def\nm#1{#1^{\left(n,m\right)}}
\global\long\def\nmpq#1{#1^{\left(n,m,p,q\right)}}
 \global\long\def\elecpushcomp#1{\mathcal{A}_{#1}}
 \global\long\def\elaselecpushcomp#1{\mathcal{B}_{#1}}
 \global\long\def\elaspushcomp#1{\mathcal{C}_{#1}}
\global\long\def\elaspushcomph#1{\hat{\mathcal{C}}_{#1}}
 \global\long\def\dnh{\aef_{DH}}
 \global\long\def\dnhc{\mu\lambda^{2}}
 \global\long\def\dnhcc{\frac{\mu}{\lambda^{2}}+\frac{1}{\matper}d_{2}^{2}}
 \global\long\def\dnhb{\frac{1}{\matper}d_{2}}
 \global\long\def\afreq{\omega}
 \global\long\def\dispot{\phi}
 \global\long\def\edpot{\varphi}
 \global\long\def\afreqh{\hat{\afreq}}
 \global\long\def\phasespeed{c}
 \global\long\def\bulkspeed{c_{B}}
 \global\long\def\speedh{\hat{c}}
 \global\long\def\dhth{\dh_{th}}
 \global\long\def\bulkspeedl{\bulkspeed_{\lambda}}
 \global\long\def\khth{\hat{k}_{th}}
 \global\long\def\p#1{#1^{\left(p\right)}}
 \global\long\def\maxinccomp#1{\inc{\outer{\sigma}}_{#1}}
 \global\long\def\maxcomp#1{\outer{\sigma}_{#1}}
 \global\long\def\relper{\matper_{r}}
 \global\long\def\sdh{\hat{d}}
 \global\long\def\iee{\varphi}
 \global\long\def\effectivemu{\tilde{\mu}}
 \global\long\def\fb#1{#1^{\left(1\right)}}
 \global\long\def\mt#1{#1^{\left(2\right)}}
 \global\long\def\phs#1{#1^{\left(p\right)}}
 \global\long\def\thc{h}
\global\long\def\nthc#1{h^{\left(#1\right)}}
 \global\long\def\state{\mathbf{s}}
 \global\long\def\harmonicper{\breve{\matper}}
 \global\long\def\kb{k_{B}}
 \global\long\def\cb{\bar{c}}
 \global\long\def\mb{\bar{\mu}}
 \global\long\def\rb{\bar{\rho}}
 \global\long\def\wavenumber{k}
 \global\long\def\rh{\hat{\mathbf{r}}}
 \global\long\def\zh{\hat{\mathbf{z}}}
 \global\long\def\th{\hat{\mathbf{\theta}}}
 \global\long\def\lz{\lambda_{z}}
 \global\long\def\lt{\lambda_{\theta}}
 \global\long\def\lr{\lambda_{r}}
 \global\long\def\st{\Omega}
 \global\long\def\stz{\Psi}
 \global\long\def\ste{\varphi}
 \global\long\def\stze{\phi}
 \global\long\def\lap{\mathcal{M}}
 \global\long\def\vh{\hat{V}}
 \global\long\def\ch{\hat{c}}
 \global\long\def\wh{\hat{\omega}}
 \global\long\def\rb{\bar{r}}
 \global\long\def\cthick{h}
 \global\long\def\vth{\Delta\vh_{th}}
 \global\long\def\kco{\kh_{co}}
 \global\long\def\normv{\Delta\hat{V}}
 \global\long\def\qh{\hat{q}_{A}}
 \global\long\def\kh{\hat{k}}
 \global\long\def\lzt{\tilde{\lambda}_{z}}
 \global\long\def\cratio{\gamma}
 \global\long\def\torusvar{\zeta}
 \global\long\def\torusfun{\mathcal{T}}
 \global\long\def\gapdensity{P}
 \global\long\def\gapdom{\mathbb{D}}
 \global\long\def\torus{\mathbb{T}^{N}}
 \global\long\def\se{\Psi}
 \global\long\def\slop{a}
\global\long\def\disp{\mathbf{u}}
\global\long\def\nh{\mathbf{n}}
\global\long\def\mh{\mathbf{m}}
 \global\long\def\pvar#1{y_{#1}}
\global\long\def\qvar#1{a_{#1}}
\global\long\def\nslop#1{a_{f}^{\left(#1\right)}}
\global\long\def\cf{\chi}
\global\long\def\gd{\varphi}
\global\long\def\ntorus#1{\mathbb{T}{}^{#1}}
\global\long\def\gap#1{\mathbb{D}^{#1}}
\global\long\def\gset{\left\{  \gamma\right\}  }
\global\long\def\nimp#1{z^{\left(#1\right)}}
\global\long\def\gbar{\bar{\gamma}}
\global\long\def\gnm#1#2{\gamma^{\left(#1,#2\right)}}
 \global\long\def\N#1{#1^{\left(N\right)}}
\global\long\def\k#1{#1^{\left(k\right)}}
\global\long\def\l#1{#1^{\left(l\right)}}
\global\long\def\zn#1{z^{\left(#1\right)}}
\global\long\def\lip{\lambda_{ip}}
\global\long\def\scomp#1#2{\sigma_{#1}^{\left(#2\right)}}
\global\long\def\fd{t_{A}}
\global\long\def\sup#1#2{#1^{\left(#2\right)}}
\global\long\def\hatgd{\hat{\gd}}
\global\long\def\hatgbar{\hat{\gbar}}
\global\long\def\hatnimp#1{\hat{z}^{\left(#1\right)}}
 \global\long\def\nc#1{c^{\left(#1\right)}}
\global\long\def\dz{\Delta\zeta}
\global\long\def\do{\Delta\omega}
\global\long\def\coeff{A}
\global\long\def\ang{\phi}
\global\long\def\angmaxcoeff{\phi_{\coeff_{c}}}
\global\long\def\angmaxdz{\phi_{\dz_{c}}}
\global\long\def\nang#1{\theta^{\left(#1\right)}}
\global\long\def\nangmax#1{\theta_{c}^{\left(#1\right)}}
\global\long\def\vtor#1{\zeta^{\left(#1\right)}}
\global\long\def\slopf{a_{f}}
\global\long\def\azi{\phi}
\global\long\def\inclin{\theta}
\global\long\def\nrho#1{\rho^{\left(#1\right)}}
\global\long\def\nmu#1{\mu^{\left(#1\right)}}
\global\long\def\fst{1^{\mathrm{st}}}
\global\long\def\njm#1{\jm^{\left(#1\right)}}

\title{On the band gap universality of\emph{ }multiphase laminates and its
applications }

\author{Ben Lustig and Gal Shmuel\thanks{Corresponding author. Tel.: +1 972 778871613. \emph{E-mail address}:
meshmuel@technion.ac.il (G. Shmuel).}\\
 {\small{}{}Faculty of Mechanical Engineering, Technion–Israel Institute
of Technology, Haifa 32000, Israel}\\
 }
\maketitle
\begin{abstract}
\citet{Shmuel2016JMPS} discovered that all infinite band structures
of waves at normal incidence in two-phase laminates are encapsulated
in a compact universal manifold. We show that manifolds of higher
dimensionality encapsulate the band structures of all multiphase laminates.
We use these manifolds to determine the density of gaps in the spectrum,
and prove it is invariant with respect to certain properties. We further
demonstrate that these manifolds are useful for formulating optimization
problems on the gaps width, for which we develop a simple bound. Using
our theory, we numerically study the dependency of the gaps density
and width on the impedance and number of phases. Finally, we show
that in certain settings, our analysis applies to non-linear multiphase
laminates, whose band diagram is tunable by finite pre-deformations.
Through simple examples, we demonstrate how the universality of our
representation is useful for characterizing this tunability in multiphase
laminates. 

\end{abstract}

\section{Introduction}

Periodicity in a transmission medium for waves renders their propagation
frequency dependent, even when the medium pointwise properties are
not \citep{Hussein2014review}. The resultant propagation can exhibit
exotic or \emph{metamaterial} characteristics, such as negative refraction
and wave steering \citep{LU2009,Zelhofer2017kochmann}. These phenomena
are not only physically intriguing, but also have functional potential
in applications such as cloaking and superlensing \citep{pendry00,milton06cloaking,Colquitt2014}.

Laminates---the media addressed in this paper---have the simplest
periodicity, as their properties change only along one direction.
Surprisingly, new results on their dynamics are still reported by
ongoing research, \emph{e.g.}, metamaterial behavior of laminates
\citep{bigoni2013prb,NematNasser2015,Willis2016jmps,Srivastava2016jmps},
field patterns in laminates with time dependent moduli \citep{Milton2017},
and dynamic homogenization of laminates \citep{NematNasser2011jmps,NN11srivastava,hanan2014prl,Srivastava2014,joseph2015}. 

We are concerned with the most familiar phenomenon: annihilation of
waves at certain frequencies and corresponding emergence of a band-gap
structure in the infinite spectrum \citep{Sigalas1992,Kushwaha1993}.
The range of these gaps varies from one laminate to another, as function
of the phase properties. For waves at normal incident angle, \citet{Shmuel2016JMPS}
discovered that all band structures of laminates with two alternating
layers are derived from a compact universal structure, \emph{independently}
of each layer thickness and specific physical properties. They used
this universality to rigorously derive the maximal width, expected
width, and density of the gaps, \emph{i.e.}, the relative width of
the gaps within the entire spectrum. Finally, \citet{Shmuel2016JMPS}
conjectured that such universality also exists for laminates made
of an arbitrary number of phases. In what follows, we provide a rigorous
proof for this conjecture, and employ it to answer several interesting
questions\emph{, e.g.}, can the gap density be increased by adding
more phases? If so, what are the optimal compositions that maximize
it? Can we enlarge specific gaps in the same way? 

The bulk of our analysis is presented in the framework of linear infinitesimal
elasticity; in the sequel we show that in certain settings it extends
to finite elasticity. Specifically, we show that our analysis applies
for incremental waves propagating in non-linear multiphase laminates
subjected to piecewise-constant finite deformations. Similarly to
the case of finitely deformed two-phase laminates, the resultant band
diagram is tunable by the static finite deformation (\citealp{Shmuel2016JMPS},
\emph{cf.}, \citealp{Zhang2017}); through simple examples, we demonstrate
how the universality of our representation is useful for characterizing
this tunability. 

We present our results in the following order. Firstly, in Sec. \ref{sec:Wave-propagation}
we revisit the derivation of the dispersion relation for multiphase
laminates, from which the band structure is evaluated. Sec. \ref{sec:A-universal-spectra}
contains our theory; therein, we show that  all infinite band structures
of multiphase laminates are encapsulated in a compact universal manifold,
whose dimensionality equals the number of layers in the periodic cell.
 We find that the gap density is the volume fraction of a universal
submanifold, derive a closed-form expression for the submanifold boundary,
and hence for the gap density. We further employ the new framework
to provide a simple bound on the gaps width and formulate corresponding
optimization problems. Sec. \ref{sec:Analysis-on-the} employs our
formulation to answer the questions posed earlier, via parametric
investigation of the compact manifold. Sec. \ref{sec:Waves-superposed-on}
details how our theory extends to non-linear multiphase laminates
of tunable band diagrams, and demonstrates its application for characterizing
this tunability. Finally, we summarize our results in Sec. \ref{sec:Concluding-remarks}. 

\section{\label{sec:Wave-propagation}Wave propagation in multiphase laminates }

\subsection{Dispersion relation}

The solution to the problem of wave propagation in periodic laminates
is well-known \citep{rytov56}; the topic receives renewed attention
recently, owing to its applications in the context of metamaterials
\citep{Willis2016jmps,Srivastava2016jmps}. For the reader convenience,
this Sec. concisely recapitulates the formulation for multiphase laminates
\citep{Lekner94}, in the framework of linear elasticity. 

A laminate is made of an infinite repetition of a unit cell comprising
$N$ layers. We denote the thickness, mass density, and Lamé coefficients
of the $n^{\mathrm{th}}$ layer by $\phase h,\phase{\rho},\phase{\lambda}$
and $\phase{\mu}$, respectively. We consider plane waves propagating
normal to the layers at frequency $\omega$. The equations of motion
are satisfied by a displacement field which can be written in each
layer as 
\begin{equation}
\phase{\disp}=\mh\left(\phase Ae^{i\phase k\nh\cdot\mathbf{x}}+\phase Be^{-i\phase k\nh\cdot\mathbf{x}}\right)e^{-i\omega t},\label{eq:uform-1}
\end{equation}
where $\nh$ is the lamination direction, $\phase A$ and $\phase B$
are constants, the wavenumber $\phase k$ equals $\frac{\omega}{\phase c}$,
$\phase c$ being the velocity that depends on the polarization $\mh$,
namely, 
\begin{equation}
\phase c=\sqrt{\frac{\phase{\tilde{\mu}}}{\phase{\rho}}}=\begin{cases}
\begin{array}{cc}
\sqrt{\frac{\phase{\mu}}{\phase{\rho}}}, & \mh\bot\nh,\\
\sqrt{\frac{\phase{\lambda}+2\phase{\mu}}{\phase{\rho}}}, & \mh\Vert\nh.
\end{array}\end{cases}\label{eq:velocity}
\end{equation}
It follows that $\phase{\disp}$ and the traction $\phase{\traction}$
at the borders of each layer are related via 
\begin{equation}
\left\{ \begin{array}{c}
\phase{\disp}\left(\mathbf{x}+\phase h\nh\right)\\
\phase{\traction}\left(\mathbf{x}+\phase h\nh\right)
\end{array}\right\} =\left[\begin{array}{cc}
\cos\p k\p h & \frac{\sin\p k\p h}{\phase{\tilde{\mu}}\p k}\\
-\phase{\tilde{\mu}}\p k\sin\p k\p h & \cos\p k\p h
\end{array}\right]\left\{ \begin{array}{c}
\phase{\disp}\left(\mathbf{x}\right)\\
\phase{\traction}\left(\mathbf{x}\right)
\end{array}\right\} .\label{eq:transfer n}
\end{equation}
 Invoking continuity conditions and sequentially applying Eq. \eqref{eq:transfer n}
yields the following relation between the fields at the ends of the
periodic cell
\begin{equation}
\left\{ \begin{array}{c}
\N{\disp}\left(\mathbf{x}+h\nh\right)\\
\N{\traction}\left(\mathbf{x}+h\nh\right)
\end{array}\right\} =\prod_{n=1}^{N}\left[\begin{array}{cc}
\cos\p k\p h & \frac{\sin\p k\p h}{\phase{\tilde{\mu}}\p k}\\
-\phase{\tilde{\mu}}\p k\sin\p k\p h & \cos\p k\p h
\end{array}\right]\left\{ \begin{array}{c}
\disp^{\left(1\right)}\left(\mathbf{x}\right)\\
\traction^{\left(1\right)}\left(\mathbf{x}\right)
\end{array}\right\} ,\label{eq:transfer N}
\end{equation}
where $h=\sum_{n=1}^{N}\n h$. Since the laminate is periodic and
linear, we also have that \citep{kitt05book_ch7,Farzbod2011}
\begin{equation}
\left\{ \begin{array}{c}
\N{\disp}\left(\mathbf{x}+h\nh\right)\\
\N{\traction}\left(\mathbf{x}+h\nh\right)
\end{array}\right\} =e^{ik_{B}h}\left\{ \begin{array}{c}
\disp^{\left(1\right)}\left(\mathbf{x}\right)\\
\traction^{\left(1\right)}\left(\mathbf{x}\right)
\end{array}\right\} ,\label{eq:bloch}
\end{equation}
where $\kb$ is the Bloch wavenumber, quantifying the fields envelope.
Eqs. (\ref{eq:transfer N}-\ref{eq:bloch}) deliver the\emph{ dispersion
relation} between $\kb$, $\omega$ and the properties of the layers,
namely,

\begin{equation}
\eta=\cos\kb h,\label{eq:dispersion-N}
\end{equation}
where\footnote{Interestingly, \citet{Kohmoto83prl} found that $\eta$ equals half
the trace of the transfer matrix in Eq. \eqref{eq:transfer N}.} 

\begin{equation}
\begin{split} 
\eta  = & \prod_{n=1}^{N}\cos\frac{\omega\n h}{\n c}-\sum_{m=1}^{N-1}\sum_{n=m+1}^{N}  \bigg[ \nm{\gamma}  \Big( \prod_{k=m,n}\sin\frac{\omega\k h}{\k c}  \Big)  \prod_{l(\neq m,n)=1}^{N}\cos\frac{\omega\l h}{\l c}  \bigg] + \\ &
+\sum_{n=1}^{N-3}\sum_{p=n+2}^{N-1}\sum_{m=n+1}^{p-1}\sum_{q=p+1}^{N}  \bigg[ \nmpq{\cratio}  \Big( \prod_{k=m,n,p,q}\sin\frac{\omega\k h}{\k c}  \Big) \prod_{l(\neq m,n,p,q)=1}^{N}\cos\frac{\omega\l h}{\l c}  \bigg]+..., 
\label{eq:eta of d-1} 
\end{split}
\end{equation} and
\[
\left\{ \nm{\gamma}=\frac{1}{2}\left(\frac{\n{\rho}\n c}{\m{\rho}\m c}+\frac{\m{\rho}\m c}{\n{\rho}\n c}\right),\!\nmpq{\gamma}=\frac{1}{2}\left(\frac{\n{\rho}\n c\p{\rho}\p c}{\m{\rho}\m c\q{\rho}\q c}+\frac{\m{\rho}\m c\q{\rho}\q c}{\n{\rho}\n c\p{\rho}\p c}\right),...\right\} \eqqcolon\left\{ \gamma\right\} 
\]
quantifies the mismatch between the impedance of the phases and their
combinations. Additional terms are added to Eq. \eqref{eq:eta of d-1}
for $N>5$. We emphasize that these terms are of a similar form to
the terms in Eq. \eqref{eq:eta of d-1}, namely, products of $\cos\frac{\omega\n h}{\n c},\,\sin\frac{\omega\n h}{\n c}$,
and impedance mismatch measures \citep{shen00}. The fact that $\eta$
maintains this functional form for arbitrary $N$ is central to our
forthcoming analysis in Sec. \ref{sec:A-universal-spectra}.

\subsection{Band structure and gap density }

\floatsetup[figure]{style=plain,subcapbesideposition=top}
\begin{figure}[t]
\centering\sidesubfloat[]{\includegraphics[width=0.3\textwidth]{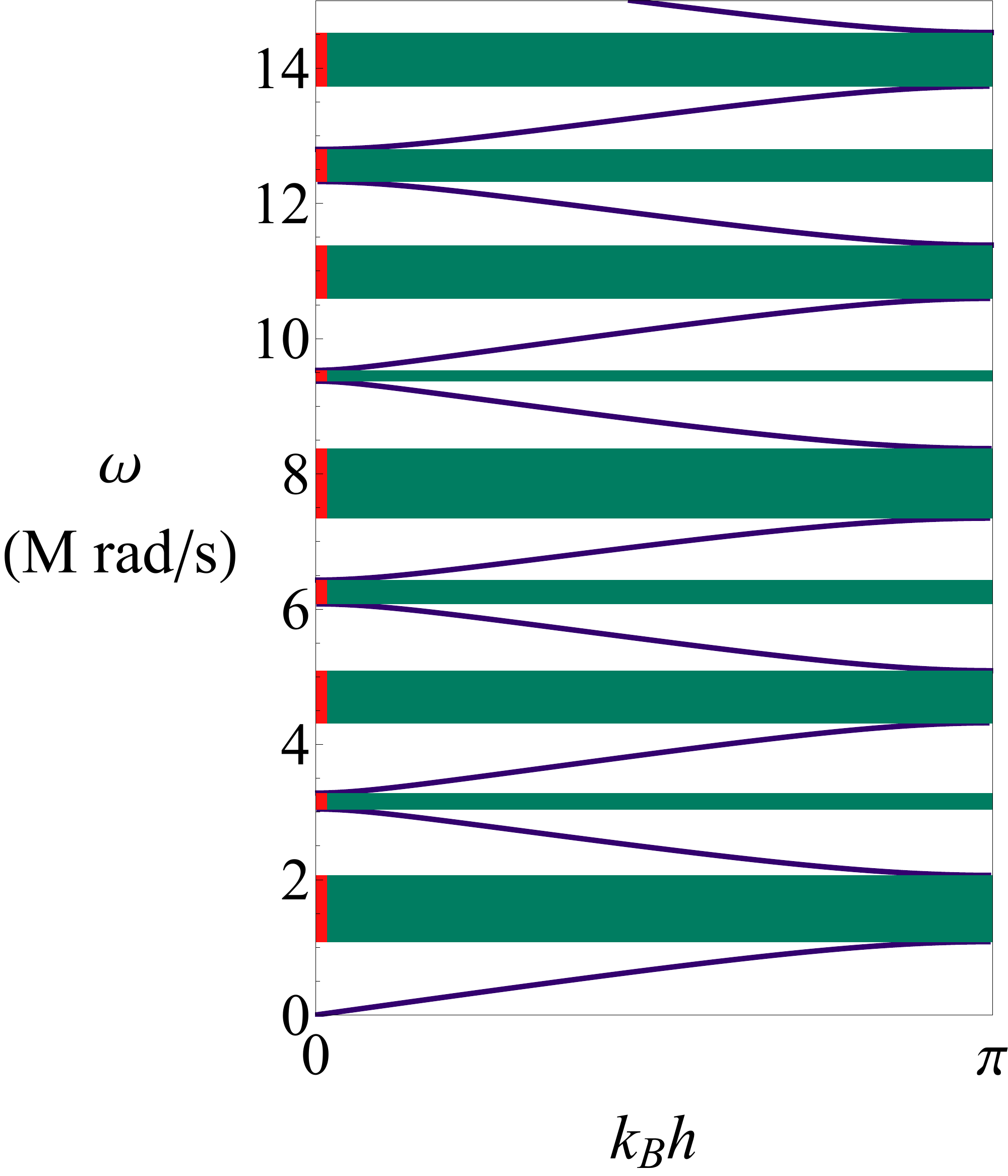}} \quad \sidesubfloat[]{\includegraphics[width=0.3\textwidth]{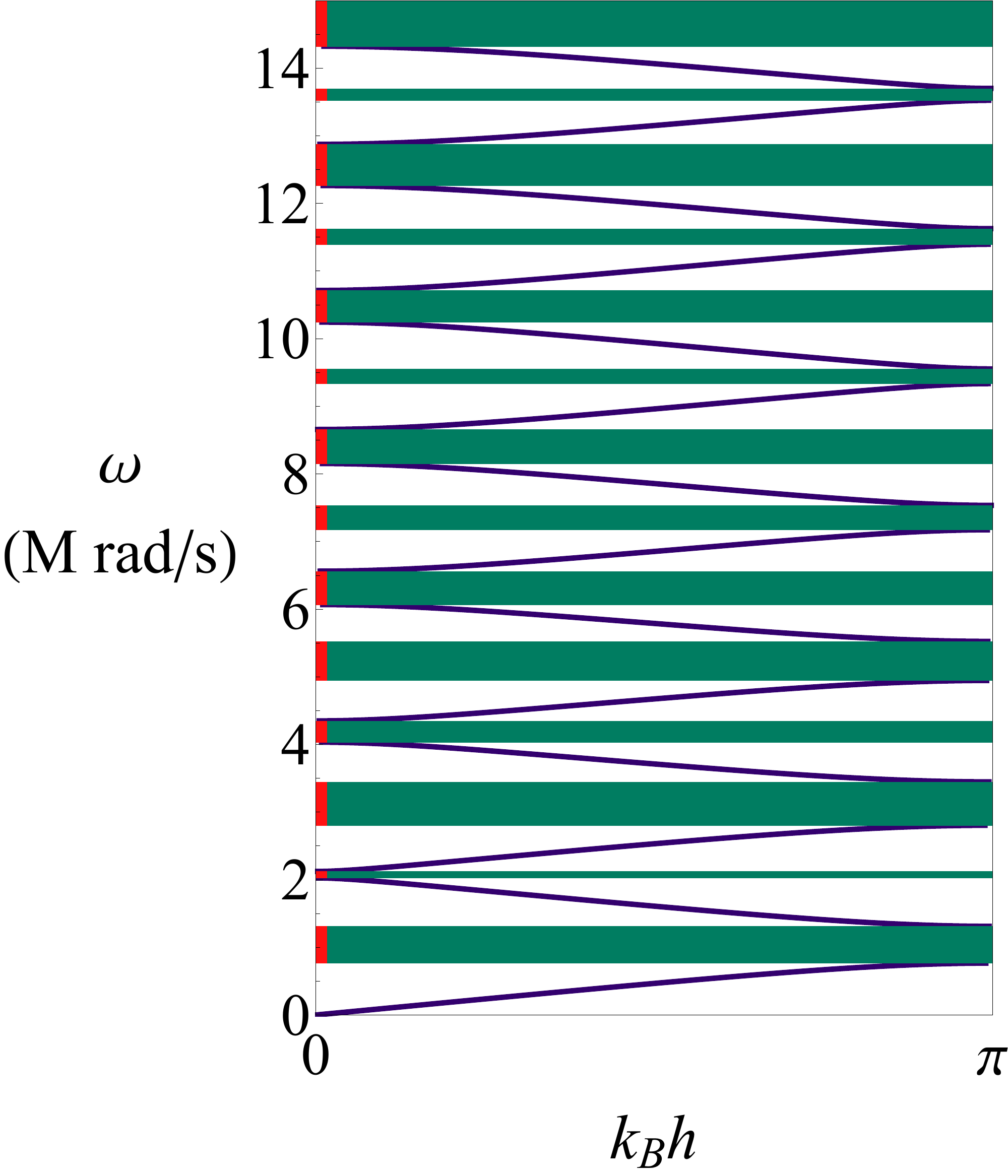}}

\bigskip

\includegraphics[width=0.8\textwidth]{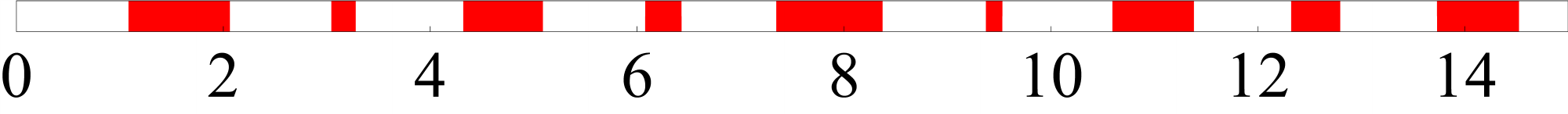}

\bigskip\bigskip

\includegraphics[width=0.8\textwidth]{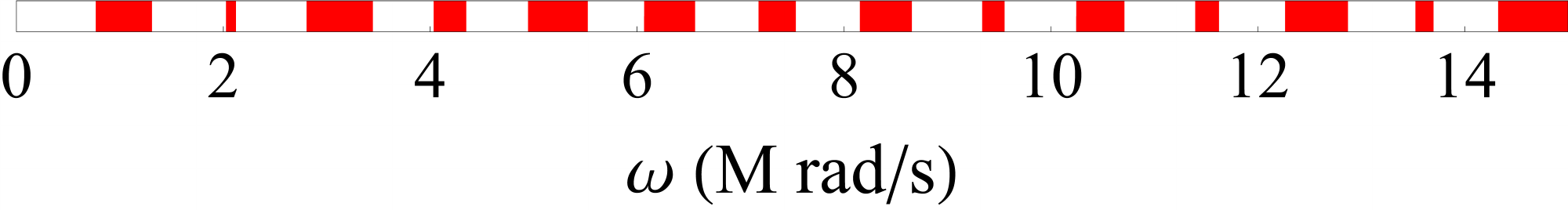}

\caption{Representative band structures of 3-layer laminates. Panels (a) and
(b) correspond to laminates \eqref{eq:laminate 1} and \eqref{eq:laminate 3},
respectively. The truncated spectra, having the range of the gaps
highlighted in red, are shown below the panels. The upper and lower
spectra correspond to panels (a) and (b), respectively. }

{\small{}{}\label{representative fig}} 
\end{figure}
Waves at a frequency $\omega$ will propagate in the laminate only
if Eq. \eqref{eq:dispersion-N} is satisfied by real $\kb$, and thus
$\left|\eta\right|<1$. Otherwise, the Bloch parameter $\kb$ is imaginary
and $\left|\eta\right|>1$, in which case the waves decay. The corresponding
frequencies create \emph{gaps} between propagating Bloch \emph{bands}
in the $\kb-\omega$ diagram, resulting in a band structure. It is
clear from Eq. \eqref{eq:dispersion-N} that the band structure depends
on the phase properties; it is also clear that the band structure
is (generally) not periodic in $\omega$. Two representative band
structures of 3-layer laminates are given in Fig. \ref{representative fig}.
Panel (a) corresponds to a laminate whose layer properties are \begin{equation}
\begin{array}{lll}
\nrho 1=2700\,\mathrm{kg/m^{3}},
& \nmu 1=26\, \mathrm{GPa}, 
& \nthc 1=3\, \mathrm{mm},\\ \nrho 2=8000\, \mathrm{kg/m^{3}}, 
& \nmu 2=80\, \mathrm{GPa}, 
& \nthc 2=2\, \mathrm{mm},\\ \nrho 3 =5000\, \mathrm{kg/m^{3}},
& \nmu 3=70\, \mathrm{GPa}, & \nthc 3=1.5\, 
\mathrm{mm}.
\end{array}
\label{eq:laminate 1}
\end{equation}The properties of layers 1 and 2 correspond to aluminum and steel,
respectively, where layer 3 corresponds to a fictitious material
whose impedance is between the impedance of layers 1 and 2. Panel
(b) corresponds to a laminate layer properties are \begin{equation}
\begin{array}{lll}
\nrho 1=4050\, \mathrm{kg/m^{3}}, & \nmu 1=39\, \mathrm{GPa}, & \nthc 1=5\, \mathrm{mm},\\
\nrho 2=12000\, \mathrm{kg/m^{3}}, & \nmu 2=120\, \mathrm{GPa}, & \nthc 2=1\, \mathrm{mm},\\
\nrho 3=7500\, \mathrm{kg/m^{3}}, & \nmu 3=105\, \mathrm{GPa}, & \nthc 3=4\, \mathrm{mm}. 
\end{array}\label{eq:laminate 3}
\end{equation} Note that the diagrams are truncated at $\omega=15\,\mathrm{M}\,\mathrm{rad/s}$;
the truncated spectra, having the range of the gaps highlighted in
red, are shown below the panels. The fact that $e^{ik_{B}h}=e^{i\left(k_{B}h+2\pi l\right)}$
for integer $l$, together with the symmetry of the $\cos$ function,
implies that identify the gaps it is sufficient to evaluate $\omega$
against $0\leq\kb h\leq\pi$.

The density of the gaps, denoted $\gd$, is their relative size in
the frequency spectrum, \emph{i.e.}, the relative length of the sum
of all red segments. We can define $\gd$ more formally as follows.
Let $\chi\left(\omega\right)$ denote the characteristic function
\begin{equation}
\chi\left(\omega\right)=\begin{cases}
1 & \omega\in\mathrm{gap},\\
0 & \mathrm{otherwise.}
\end{cases}\label{eq:characteristic function}
\end{equation}
In terms of $\chi,$ the relative size of the gaps in a frequency
range $\left[0,\omega_{0}\right]$ is 
\begin{equation}
\gd_{\omega_{0}}=\frac{1}{\omega_{0}}\int_{0}^{\omega_{0}}\cf\left(\omega\right)\mathrm{d}\omega,\label{eq:sequence gd}
\end{equation}
and the gap density is 

\begin{equation}
\gd\coloneqq\lim_{\omega_{0}\rightarrow\infty}\gd_{\omega_{0}}.\label{eq:gd definition}
\end{equation}

\noindent Only recently, \citet{Shmuel2016JMPS} proved that for laminates
made of two alternating layers the above limit exists, and provided
an integral of a closed-form expression for its value, thus were able
to calculate the gap density. The case of multilayer periodic cells
case is next. 

\section{\label{sec:A-universal-spectra}Universal gap structure in a compact
manifold }

\citet{Shmuel2016JMPS} found that all infinite band structures of
two-layer laminates are encapsulated in a compact universal manifold,
independent of the thickness of each layer and its specific physical
properties; they further conjectured that this encapsulation can be
generalized to an arbitrary number of layers. In the first part of
this Sec., we prove this conjecture, derive a closed-form expression
for the manifold that encapsulates the gap structure of multiphase
laminates, and in turn, determine the corresponding gap density in
terms of the manifold volume.

\citet{Shmuel2016JMPS} further recognized that the compact structure
is also useful for formulating optimization problems on the gaps width,
and derived simple bounds 2-layer laminates. In the second part of
this Sec., we generalize their approach to multiphase laminates.  

\subsection{Morphism between spectral band gaps and the \emph{N}-dimensional
torus}

Following the approach in \citet{Shmuel2016JMPS}\footnote{This was developed from a method in quantum graphs analysis \citep{BarGas_jsp00,band13prl}. },
we begin with the following substitution of variables 
\begin{equation}
\n{\torusvar}\coloneqq\frac{\omega\n h}{\n c},\label{eq:substitution of variables}
\end{equation}
which renders $\eta$ a $2\pi$-periodic function in each one of the
new variables $\n{\torusvar}$,\emph{ i.e.}, 
\begin{equation}
\eta\left(\torusvar^{\left(1\right)},\torusvar^{\left(2\right)},...,\torusvar^{\left(n\right)}+2\pi,...\torusvar^{\left(N\right)};\left\{ \gamma\right\} \right)=\eta\left(\torusvar^{\left(1\right)},\torusvar^{\left(2\right)},...,\torusvar^{\left(n\right)},...\torusvar^{\left(N\right)};\left\{ \gamma\right\} \right).\label{eq:eta periodicity}
\end{equation}
This implies that the domain of $\eta$ can be represented by an \emph{N}-dimensional
torus whose coordinates are $\left(\torusvar^{\left(1\right)},\torusvar^{\left(2\right)},...,\torusvar^{\left(N\right)}\right)$,
such that each coordinate is defined modulo $2\pi$. The torus is
equivalent to an \emph{N}-dimensional cube, whose opposite sides are
identified. In what follows, we interchange between the terms torus
and cube, with the understanding that they are equivalent. The function
$\eta$ linearly depends on $\sin\n{\torusvar}$ and $\cos\n{\zeta}$,
which change sign under the transformation $\n{\torusvar}\rightarrow\n{\torusvar}+\pi$,
and hence $\eta$ changes sign too. It follows that $\left|\eta\right|$
is $\pi$-periodic in $\n{\torusvar}$, hence defined over an \emph{N}-dimensional
torus, or \emph{N}-cube, whose edges are of length $\pi$. Since the
existence condition for Bloch waves depends on the absolute value
of $\eta$, in what follows we focus on $\left|\eta\right|$ and the
$\pi$-periodic torus. On this $\pi$-periodic torus, denoted $\ntorus N$,
the equations $\frac{\mathrm{d}\n{\torusvar}}{\mathrm{d}\omega\ ^{{\color{white}n}}}=\frac{\n h}{\n c}$
define a linear flow $\left\{ \overrightarrow{\torusvar}\right\} _{\omega\in\mathbb{R}}$in
the form 
\begin{equation}
\overrightarrow{\torusvar}\left(\omega\right)=\left(\frac{h^{\left(1\right)}}{c^{\left(1\right)}},\frac{h^{\left(2\right)}}{c^{\left(2\right)}},...,\frac{h^{\left(N\right)}}{c^{\left(N\right)}}\right)\omega\ \mathrm{mod}\ \pi.\label{eq:zeta linear flow}
\end{equation}
The flow propagates uniformly in the \emph{N}-cube as a line; we arbitrarily
interpret $\torusvar^{\left(N\right)}$ as the flow height, such that
its slope in each one of the coordinates $\torusvar^{\left(n\right)}$
is 
\begin{equation}
\nslop n\coloneqq\frac{\torusvar^{\left(N\right)}}{\torusvar^{\left(n\right)}}=\frac{h^{\left(N\right)}}{c^{\left(N\right)}}\frac{\n c}{\n h}.\label{eq:slop}
\end{equation}
When the flow line reaches a cube face, it continues at the opposite
one, owing to their identification. In the supplementary videos (available
online), we present the morphism between the frequency axis in the
band structure and the flow evolution on $\ntorus 3$, for two laminates
of different microstructure, and identical phased physical properties,
given by Eq. \eqref{eq:laminate 1}. 

If the slopes $\left\{ \nslop n\right\} $ are rationally independent,
\emph{viz.}, any set of integers $\left\{ \alpha^{\left(n\right)}\right\} $
satisfies 
\begin{equation}
\sum_{n=1}^{N}\alpha^{\left(n\right)}\nslop n\neq0,\label{eq:irrationally independent}
\end{equation}
except the trivial set $\alpha^{\left(n\right)}=0$, then the flow
covers the torus, as depicted by the supplementary videos. Furthermore,
it implies that the linear flow is ergodic, such that a\emph{verages
over $\omega$ in the frequency domain are equivalent to averages
over the torus}\footnote{For complete details on the corresponding theorems, we refer to the
excellent treatise by \citet{katok96book}.}.  We are specifically interested in the gap density $\gd$---the
average of $\cf$, for which ergodicity implies 
\begin{equation}
\varphi=\frac{1}{V}\int_{\ntorus N}\cf\left(\overrightarrow{\torusvar}\right)\mathrm{d}V,\label{eq:torus integral-1}
\end{equation}
where $\mathrm{d}V=\mathrm{d}\torusvar^{\left(1\right)}\mathrm{d}\torusvar^{\left(2\right)}...\mathrm{d}\torusvar^{\left(N\right)}$.
We recall that $\cf$ equals 1 if a certain frequency belongs to a
gap, which on the torus translates to the condition $\left|\eta\left(\overrightarrow{\torusvar}\right)\right|>1$.
In other words, gaps are identified with the intersection of the flow
with a subset of the torus whose image is greater than 1; we denote
this subset by $\gap N$. Hence,\emph{ the gap density equals to the
relative volume of this subset in the torus}, namely, 
\begin{equation}
\gd=\frac{\mathrm{vol}\gap N}{\mathrm{vol}\ntorus N}.\label{eq:volume ratio on torus}
\end{equation}
  We refer again to the supplementary videos, where $\gap 3$ is
depicted in green; since the laminates considered have the same $\left\{ \gamma\right\} $,
the gaps of the two flows are derived from the same $\gap 3$; since
the laminates microstructure is different, their flows differ in their
direction.

For bilayer laminates, \citet{Shmuel2016JMPS} calculated the relative
volume (area) by deriving a closed-form expression for the envelope
(curve) of $\mathbb{D}^{2}$; we provide next a simple procedure to
derive a closed-form expression for envelope of $\gap N$ for any
$N$, \emph{i.e.}, when the unit cell comprises an arbitrary number
of $N$ layers. The description of these hypersurfaces---on which
\begin{equation}
\left|\eta\right|=1,\label{eq:eta one}
\end{equation}
requires an expression for $\zeta^{\left(N\right)}$ in terms of $\left\{ \torusvar^{\left(1\right)},...,\torusvar^{\left(N-1\right)}\right\} $.
To derive such expression, we define the variables 
\begin{equation}
\pvar n\coloneqq\tan\frac{\n{\zeta}}{2},\label{eq:polynim variables}
\end{equation}
and observe that in terms of $\left\{ \pvar n\right\} $, the harmonic
functions comprising $\left|\eta\right|$ are $\cos\n{\torusvar}=\frac{1-\pvar n^{2}}{1+\pvar n^{2}}$
and $\sin\n{\torusvar}=\frac{2\pvar n}{1+\pvar n^{2}}$. We multiply
Eq. \eqref{eq:eta one} by $1+\pvar N^{2}$, to obtain a quadratic
equation for $\pvar N$, whose constants are combinations of $\left\{ \gamma\right\} $,
$\left(\pvar 1,...,\pvar{N-1}\right)$ and $\left(\pvar 1^{2},...,\pvar{N-1}^{2}\right)$.
Finally, by inverting Eq. \eqref{eq:polynim variables} for $n=N$
and substituting the solution for $\pvar N$, we achieve a closed-form
expression for $\zeta^{\left(N\right)}$. Accordingly for $\eta=-1$,
we have that 
\begin{equation}
\torusvar^{\left(N\right)}=2\arctan\frac{-\qvar 1\pm\sqrt{\qvar 1^{2}-4\qvar 2\qvar 0}}{2\qvar 2},\label{eq:zeta N}
\end{equation}
whose constants are 
\begin{equation}
\qvar 0=1,\ \ \qvar 1=-2\gamma^{\left(1,2\right)}\tan\frac{\vtor 1}{2},\ \ \qvar 2=\tan^{2}\frac{\vtor 1}{2},\label{eq:N2}
\end{equation}
when $N=2$; when $N=3$, the constants are 

\begin{equation} 
\begin{split}
& \qvar 0=\tan^{2}\frac{\vtor 1}{2}\tan^{2}\frac{\vtor 2}{2}-2\gnm 12\tan\frac{\vtor 1}{2}\tan\frac{\vtor 2}{2}+1,  \\
& \qvar 1=2\gnm 13\tan\frac{\vtor 1}{2}\left(\tan^{2}\frac{\vtor 2}{2}-1\right)+2\gnm 23\tan\frac{\vtor 2}{2}\left(\tan^{2}\frac{\vtor 1}{2}-1\right), \\
& \qvar 2=\tan^{2}\frac{\vtor 1}{2}+\tan^{2}\frac{\vtor 2}{2}+2\gnm 12\tan\frac{\vtor 1}{2}\tan\frac{\vtor 2}{2}, \\
\end{split}
\label{eq:N3} 
\end{equation}and so on, for any $N$. 

We conclude this part with remarks on laminates whose slopes $\left\{ \nslop n\right\} $
are not rationally independent. The set of such laminates constitutes
a subset with measure zero on the set of all possible laminates, since
there are uncountably many rationally independent sets $\left\{ \nslop n\right\} $,
and countably many rationally dependent sets $\left\{ \nslop n\right\} $.
Therefore, this is a negligible set which corresponds to an irregular
case, contrary to the generic case analyzed above, when the flow uniformly
covers the whole torus. The only case in which the band diagram is
periodic, and therefore also the orbit of the flow, is when all the
slopes are rational numbers. Clearly, in this case the gap density
is different than \eqref{eq:volume ratio on torus}, and equals the
relative length of the flow intersection with $\gap{}$ over the length
of one orbit. When $N=2$, the flow is either periodic or ergodic;
when $N>2$, there exists the following additional scenario. Say only
$\nslop m$ is rational and the remaining slopes are rationally independent.
Then, the flow remains only on hyperplanes in the cube that are defined
by $\torusvar^{\left(N\right)}=\torusvar^{\left(m\right)}\nslop m$,
and covers uniformly only these hyperplanes. Similarly, if another
slope is also rational, say $\nslop l$, then the flow remains on
the intersection of hyperplanes defined by $\torusvar^{\left(N\right)}=\torusvar^{\left(m\right)}\nslop m$
and $\torusvar^{\left(N\right)}=\torusvar^{\left(l\right)}\nslop l$,
covering it uniformly. This reduction of the flow orbit continues
if additional slopes are rational, until the periodic case is obtained
when all the slopes are rational numbers.

\subsection{\label{subsec:Gap-width-analysis}Gap width analysis on $\protect\ntorus N$}

Potential applications such as noise filters and vibration isolators
require wide gaps; topology optimization aims at tailoring the medium
microstructure to this objective \citep{Sigmund2003,Bilal2011PRE,bortot2018b}.
In this Sec., we employ our framework to formulate in a simple way
these optimization problems on the compact manifold $\gap N$, and
as a byproduct, derive a simple bound on the gaps width. 

Our starting point is the identification of the gaps width on $\ntorus N$.
Say the flow enters $\gap N$ at $\overrightarrow{\torusvar}\left(\omega_{l}\right)=\left(\frac{h^{\left(1\right)}}{c^{\left(1\right)}},\frac{h^{\left(2\right)}}{c^{\left(2\right)}},...,\frac{h^{\left(N\right)}}{c^{\left(N\right)}}\right)\omega_{l}$,
and exits at $\overrightarrow{\torusvar}\left(\omega_{u}\right)=\left(\frac{h^{\left(1\right)}}{c^{\left(1\right)}},\frac{h^{\left(2\right)}}{c^{\left(2\right)}},...,\frac{h^{\left(N\right)}}{c^{\left(N\right)}}\right)\omega_{u}$.
The distance between the two coordinates over the torus is 
\begin{equation}
\dz\coloneqq\left|\overrightarrow{\torusvar}\left(\omega_{u}\right)-\overrightarrow{\torusvar}\left(\omega_{l}\right)\right|=\do\sqrt{\sum_{n=1}^{N}\frac{h^{\left(n\right)^{2}}}{c^{\left(n\right)^{2}}}},\label{eq:dz dw}
\end{equation}
where $\do\coloneqq\omega_{u}-\omega_{l}$ is the gap width. (See
Fig. \ref{3layers}(b) for an exemplary $\dz$ segment of $\gap 3$.)
It follows that maximizing $\do$ over an admissible set of unit cell
compositions, denoted $\mathcal{S}\left(h^{\left(n\right)},c^{\left(n\right)}\right)$,
is equivalent to maximizing $\dz\left(\sum_{n=1}^{N}\frac{h^{\left(n\right)^{2}}}{c^{\left(n\right)^{2}}}\right)^{-\frac{1}{2}}$
over its corresponding admissible set. The merit in the expression
that contains $\dz$ is threefold. (\emph{i}) It is formulated over
the torus, hence encapsulates the whole spectra. (\emph{ii}) Since
we derived a closed-form expression for $\torusvar^{\left(N\right)}$
in Eq. \eqref{eq:zeta linear flow}, at times a closed-form expression
for $\dz$ is accessible too. (\emph{iii}) Eq. \eqref{eq:dz dw} lends
itself for the basis of the following simple bound 
\begin{equation}
\max_{\mathcal{S}\left(h^{\left(n\right)},c^{\left(n\right)}\right)}\do=\max_{\mathcal{S}\left(\nslop n,\overrightarrow{b}\right)}\left\{ \dz\left(\sum_{n=1}^{N}\frac{h^{\left(n\right)^{2}}}{c^{\left(n\right)^{2}}}\right)^{-\frac{1}{2}}\right\} \leq\max_{\mathcal{S}\left(\nslop n,\overrightarrow{b}\right)}\left\{ \dz\right\} \,\max_{\mathcal{\mathcal{S}}\left(h^{\left(n\right)},c^{\left(n\right)}\right)}\left\{ \left(\sum_{n=1}^{N}\frac{h^{\left(n\right)^{2}}}{c^{\left(n\right)^{2}}}\right)^{-\frac{1}{2}}\right\} ,\label{eq:do dz}
\end{equation}
where $\mathcal{S}\left(\nslop n,\overrightarrow{b}\right)$ is the
corresponding set of admissible slops $\nslop n$ and intersection
points $\overrightarrow{b}$ with the cube faces that characterize
the linear flow lines. (Fig. \ref{3layers}(b) depicts such exemplary
point $\overrightarrow{b}=\left(0,2,1.3\right)$ in $\ntorus 3$.)
In the sequel (Sec. \ref{subsec:Applications-in-gap}), this analysis
will be made more concrete via a specific optimization problem. 

\section{\label{sec:Analysis-on-the}Parametric investigation on $\protect\ntorus N$}

\subsection{\label{subsec:3-layer-laminates}Gap density analysis}

\floatsetup[figure]{style=plain,subcapbesideposition=top}

\begin{figure}[t]
\centering\sidesubfloat[]{\includegraphics[width=0.3\textwidth]{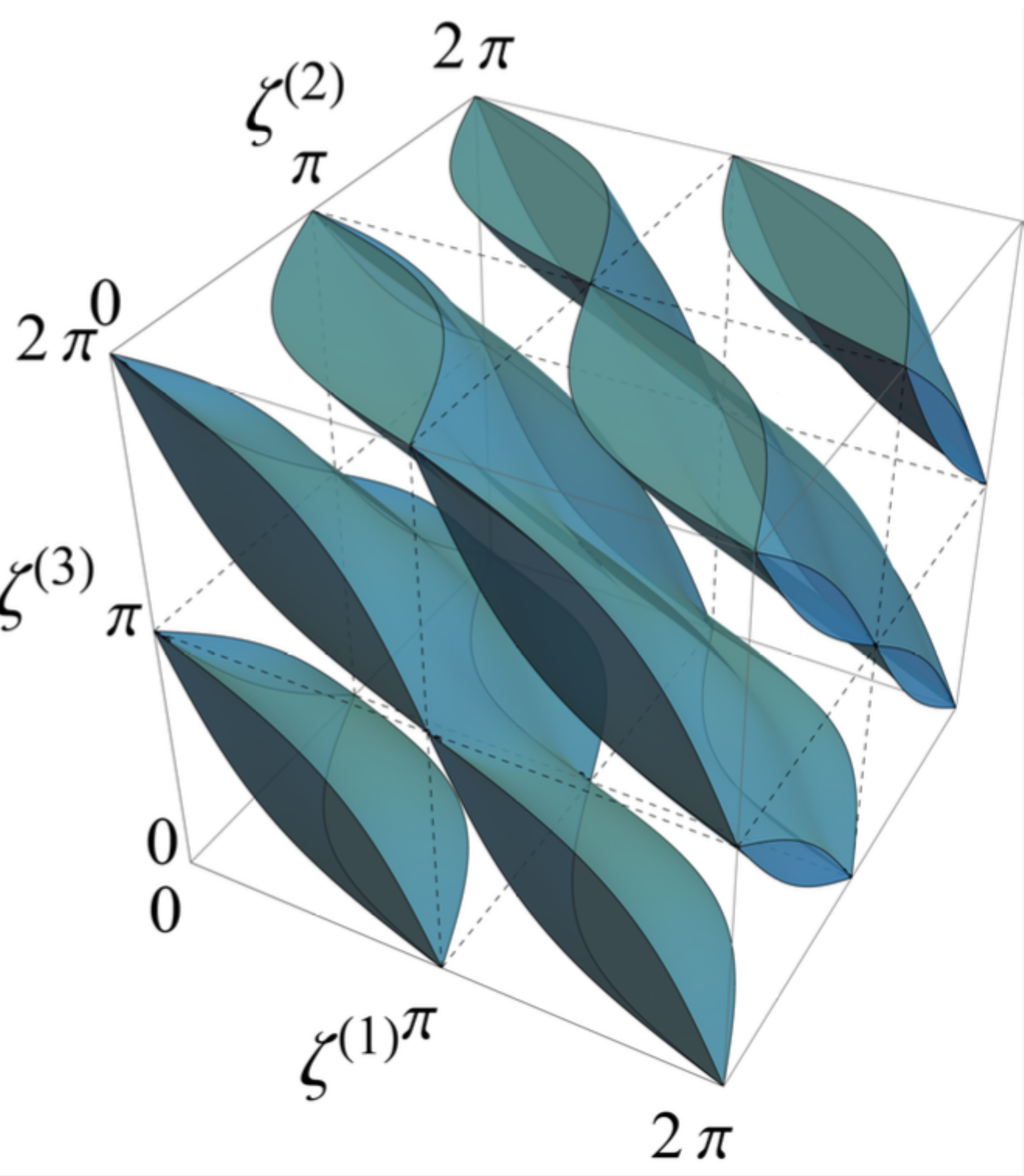}}  \sidesubfloat[]{\includegraphics[width=0.3\textwidth]{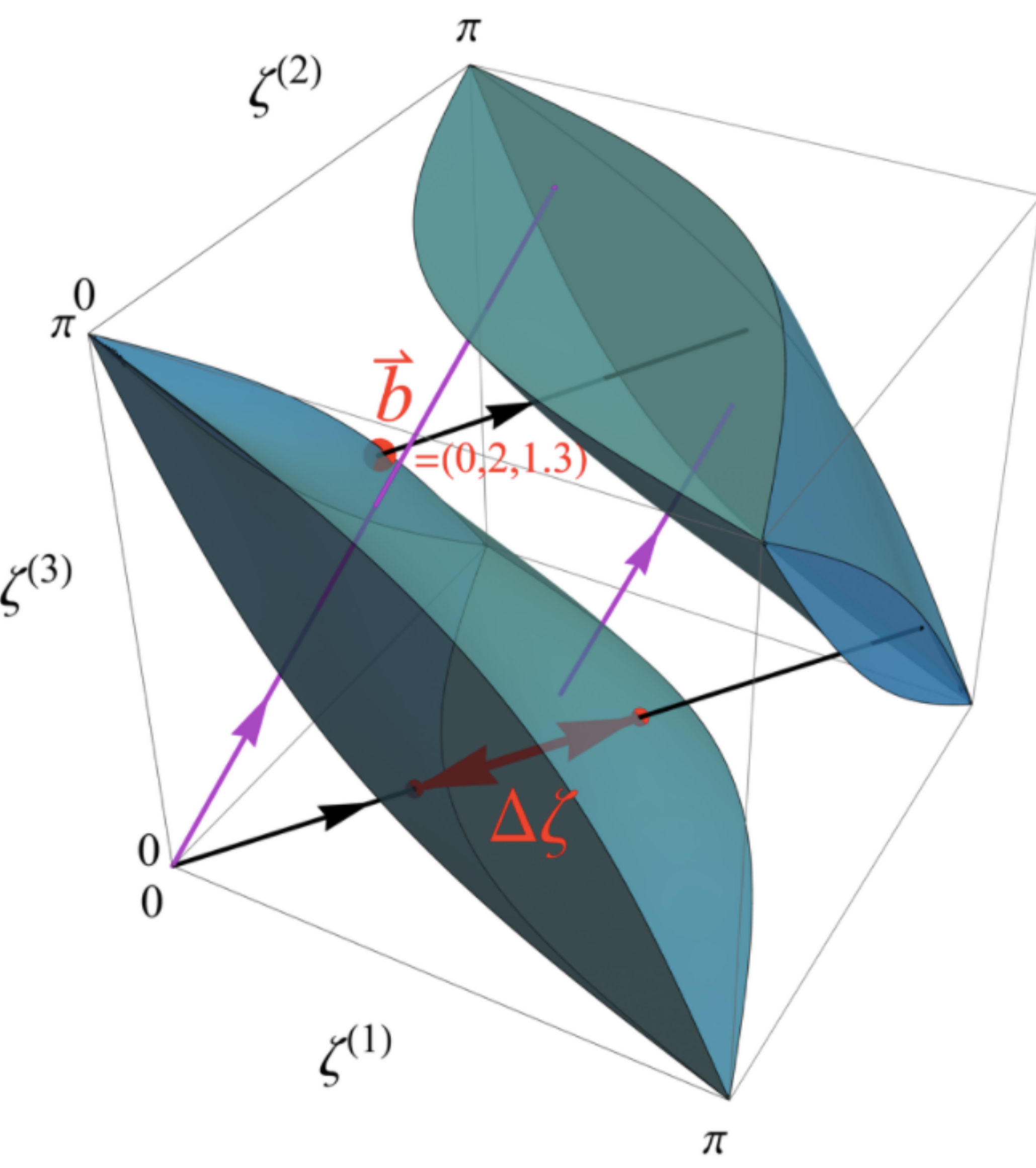}}\sidesubfloat[]{\includegraphics[width=0.3\textwidth]{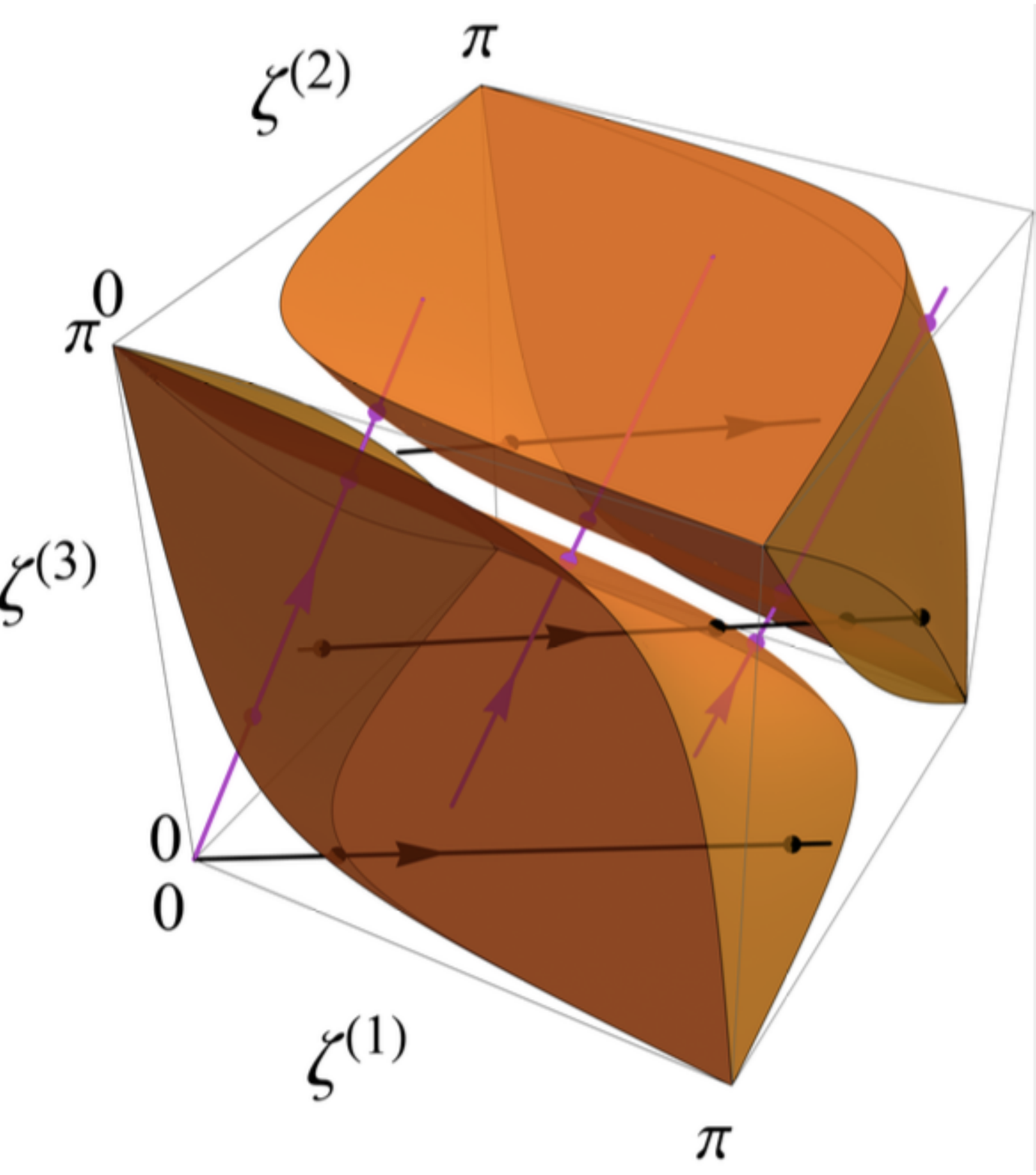}}\caption{The envelope of $\protect\gap 3$ in the (a) 2$\pi$-periodic and
(b) $\pi$-periodic cubes, of laminates with $\left\{ \gamma\right\} =\{1.675,\ 1.34,\ 1.045\}$.
Representative flow lines of laminates \eqref{eq:laminate 1} and
\eqref{eq:laminate 2} are depicted in black and purple, respectively.
(c) $\pi$-periodic envelope of $\protect\gap 3$ for laminates with
$\left\{ \gamma\right\} =\{6.76,\ 5.021,\ 1.045\}$. Part of two flows,
when phase $1$ of laminate \eqref{eq:laminate 1} is replaced by
PMMA given in Eq. \eqref{eq:constit laminate 4 and 5}, with microstructures
as in Eq. \eqref{eq:thickness laminate 4 and 5}, are depicted in
black and purple, respectively.}

{\small{}{}\label{3layers}} 
\end{figure}
\emph{3-layer laminates}. We begin with an analysis of 3-layer laminates,
whose torus dimensionality enables its illustration. Fig. \ref{3layers}(a)
shows the envelope of $\gap 3$ in the 2$\pi$-cube of laminates with
$\left\{ \gamma\right\} \coloneqq\{\gnm 12,\ \gnm 13,\ \gnm 23\}=\{1.675,\ 1.34,\ 1.045\}$,
\emph{e.g.}, laminate \eqref{eq:laminate 1}. Indeed, the structure
is $\pi$-periodic, and the reduced $\pi$-periodic cube is depicted
in Fig. \ref{3layers}(b). Therein, we also plot in black a part of
the flow that corresponds to laminate \eqref{eq:laminate 1}. For
comparison, a partial flow of a laminate with the same constituents,
however with different thicknesses, namely, 

\begin{equation} 
\begin{array}{lll}
\nthc 1=1\, \mathrm{mm}, & \nthc 2=2\, \mathrm{mm}, & \nthc 3=3.5\, \mathrm{mm},
\end{array}
\label{eq:laminate 2}
\end{equation}is depicted in purple. 

We illustrate the effect $\left\{ \gamma\right\} $ has on the domain
of $\gap 3$ by replacing the first constituent of laminate \eqref{eq:laminate 1}
with PMMA, whose physical properties are

\begin{equation} 
\begin{array}{ll}
\nrho 1=1180\, \mathrm{kg/m^{3}}, & \nmu 1=3\, \mathrm{GPa},
\end{array}
\label{eq:constit laminate 4 and 5}
\end{equation}

\noindent and evaluating the resultant envelope in Fig. \ref{3layers}(c),
which corresponds to $\left\{ \gamma\right\} =\{6.76,\ 5.021,\ 1.045\}$.
Parts of two representative flows, associated with the microstructures 

 \begin{equation} 
\begin{array}{lll}
\nthc 1=3\, \mathrm{mm}, & \nthc 2=2\, \mathrm{mm}, & \nthc 3=1.5\, \mathrm{mm},\\
\nthc 1=1\, \mathrm{mm}, & \nthc 2=2\, \mathrm{mm}, & \nthc 3=7.5\, \mathrm{mm},\\
\end{array}
\label{eq:thickness laminate 4 and 5}
\end{equation}

\noindent are depicted in black and purple, respectively. 

\begin{figure}[t]
\centering\includegraphics[width=0.8\textwidth]{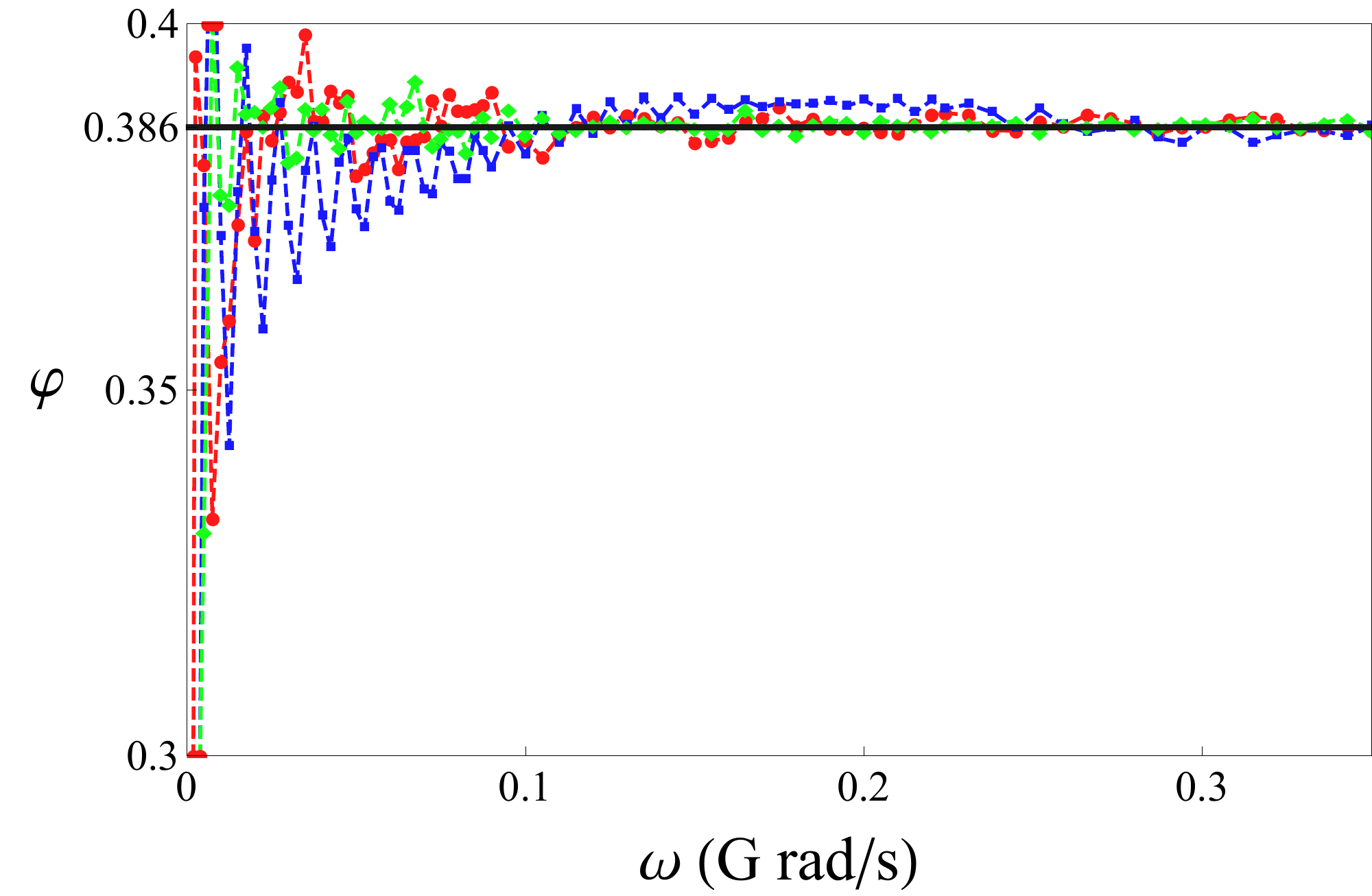}\caption{The gap density of laminates \eqref{eq:laminate 1}, \eqref{eq:laminate 3}
and \eqref{eq:laminate 2}, calculated on $\protect\ntorus 3$ (black
line), and corresponding sequences $\varphi_{\omega_{0}}$, calculated
directly from the original band structure (red and blue and green
marks, respectively). }

{\small{}{}\label{convergence GD}} 
\end{figure}
Fig. \ref{convergence GD} compares the calculation of the gap density
via Eq. \eqref{eq:volume ratio on torus} on $\ntorus 3$ (black line),
with sequences $\varphi_{\omega_{0}}$, calculated directly from the
original band structure at increasing values of $\omega_{0}$. Specifically,
we calculate the sequence for laminates \eqref{eq:laminate 1} and
\eqref{eq:laminate 2}, and denote its values at discrete $\omega_{0}$
by the red and blue marks, respectively. A third sequence (green marks)
corresponds to laminate \eqref{eq:laminate 3} whose properties are
different from laminates \eqref{eq:laminate 1} and \eqref{eq:laminate 2},
yet yield the same $\gset$. Indeed, the sequences $\varphi_{\omega_{0}}$
converge to $\frac{\mathrm{vol}\gap 3}{\mathrm{vol}\ntorus 3}$, as
our theory predicts.

\floatsetup[figure]{style=plain,subcapbesideposition=top}

\begin{figure}[t]
\centering\sidesubfloat[]{\includegraphics[width=0.45\textwidth]{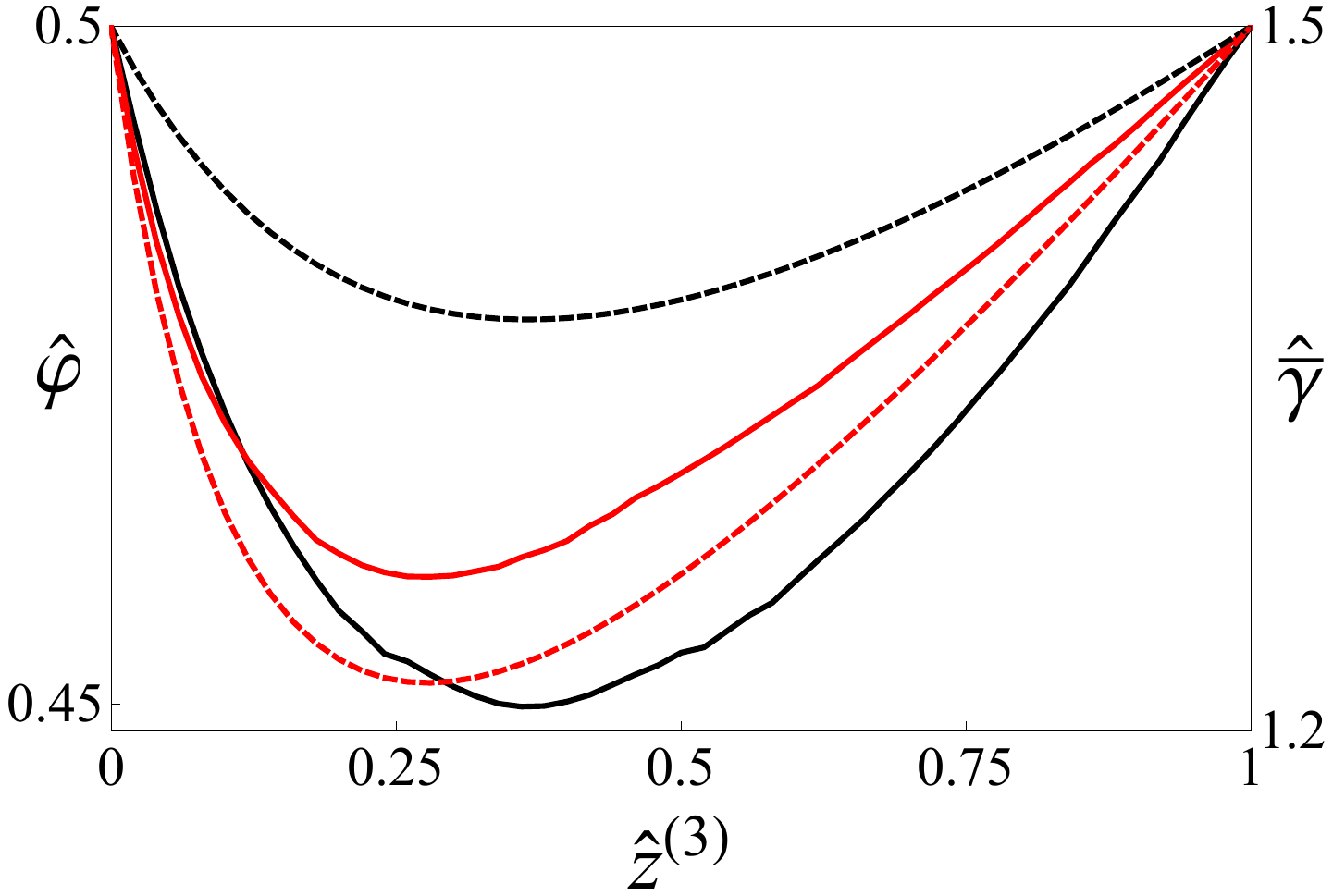}} \quad \sidesubfloat[]{\includegraphics[width=0.45\textwidth]{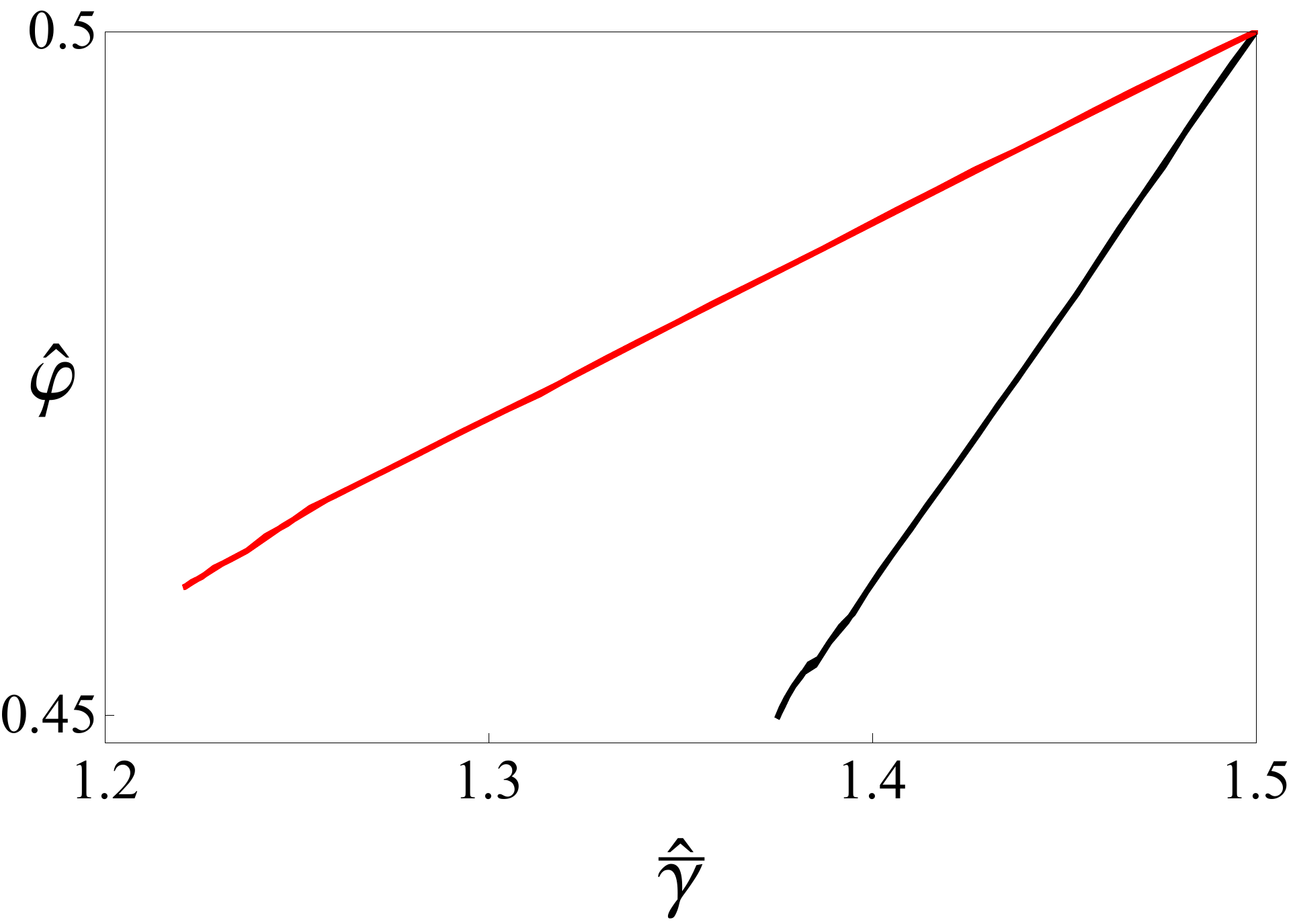}}\caption{(a) Normalized gap density $\hat{\protect\gd}$ (continuous curves)
and $\hat{\protect\gbar}$ (dashed curves) as functions of phase 3
normalized impedance. Black curves correspond to a laminate whose
phases 1 and 2 impedance is as in \eqref{eq:laminate 1}. Red curves
correspond to a laminate whose phase 1 (resp. 2) impedance is $2/3$
(resp. $3/2$) the impedance of phase 1 (resp. 2) in \eqref{eq:laminate 1}.
(b) Normalized gap density as function of $\hat{\protect\gbar}$ . }

{\small{}{}\label{GD of z3}} 
\end{figure}
The result that the gap density for classes of $\gset$ is universal
and computable via Eq. \eqref{eq:torus integral-1} is exploited next
to investigate if we can increase the gap density of laminates of
two alternating layers by introducing a third phase. To this end,
in Fig. \ref{GD of z3} we carry out a parametric investigation of
the relation between the gap density, $\left\{ \gamma\right\} $ and
the impedance of the third phase, for a laminate whose phases 1 and
2 are as of laminate \eqref{eq:laminate 1}. Panel \ref{GD of z3}(a)
shows the gap density (continuous black curves) and $\gbar=\frac{\gnm 12+\gnm 13+\gnm 23}{3}$
(dashed black curves) as functions of phase 3 impedance, normalized
according to \begin{equation}
\begin{split}
& \hatnimp 3\coloneqq\frac{\nimp 3-\nimp 1}{\nimp 2-\nimp 1}, \\
& \hatgd\coloneqq0.5\frac{\gd}{\gd(\nimp 3=\nimp 1)}, \\
& \hatgbar\coloneqq1.5\frac{\gbar}{\gbar(\nimp 3=\nimp 1)}.
\end{split}
\label{eq:normalized}
\end{equation}We observe that $\gbar$ is smaller than its value for the bilayer
laminate if the impedance of phase 3 is between the impedances of
phases 1 and 2, and the value of the gap density is lower too. Conversely,
if the impedance of phase 3 is greater (resp. lower) the the impedance
of the phase 2 (resp. 1), the gap density is higher (not shown in
the figure). We find that this trend is independent of the specific
values of the impedance of phases 1 and 2. For example, we consider
a laminate whose phase 1 (resp. 2) impedance is $2/3$ (resp. $3/2$)
the impedance of phase 1 (resp. 2) in laminate \eqref{eq:laminate 1}.
Its normalized gap density $\hat{\gd}$ (red continuous curves) and
$\hat{\gbar}$ (red dashed curves) as functions of phase 3 impedance
demonstrate the same trend. A different representation of this dependency
is given in Fig. \ref{GD of z3}(b), in which the gap density is depicted
versus $\hat{\gbar}$. Indeed, we observe that the gap density is
a monotonically increasing function of $\hat{\gbar}$. We conclude
that when subjected to the constraint 
\begin{equation}
\nimp 1<\nimp 3<\nimp 2,\label{eq:z1z3z2}
\end{equation}
the gap density of 3-layer laminates cannot exceed the gap density
of  2-layer laminates. 

\floatsetup[figure]{style=plain,subcapbesideposition=top}

\begin{figure}[t]
\centering\sidesubfloat[]{\includegraphics[width=0.45\textwidth]{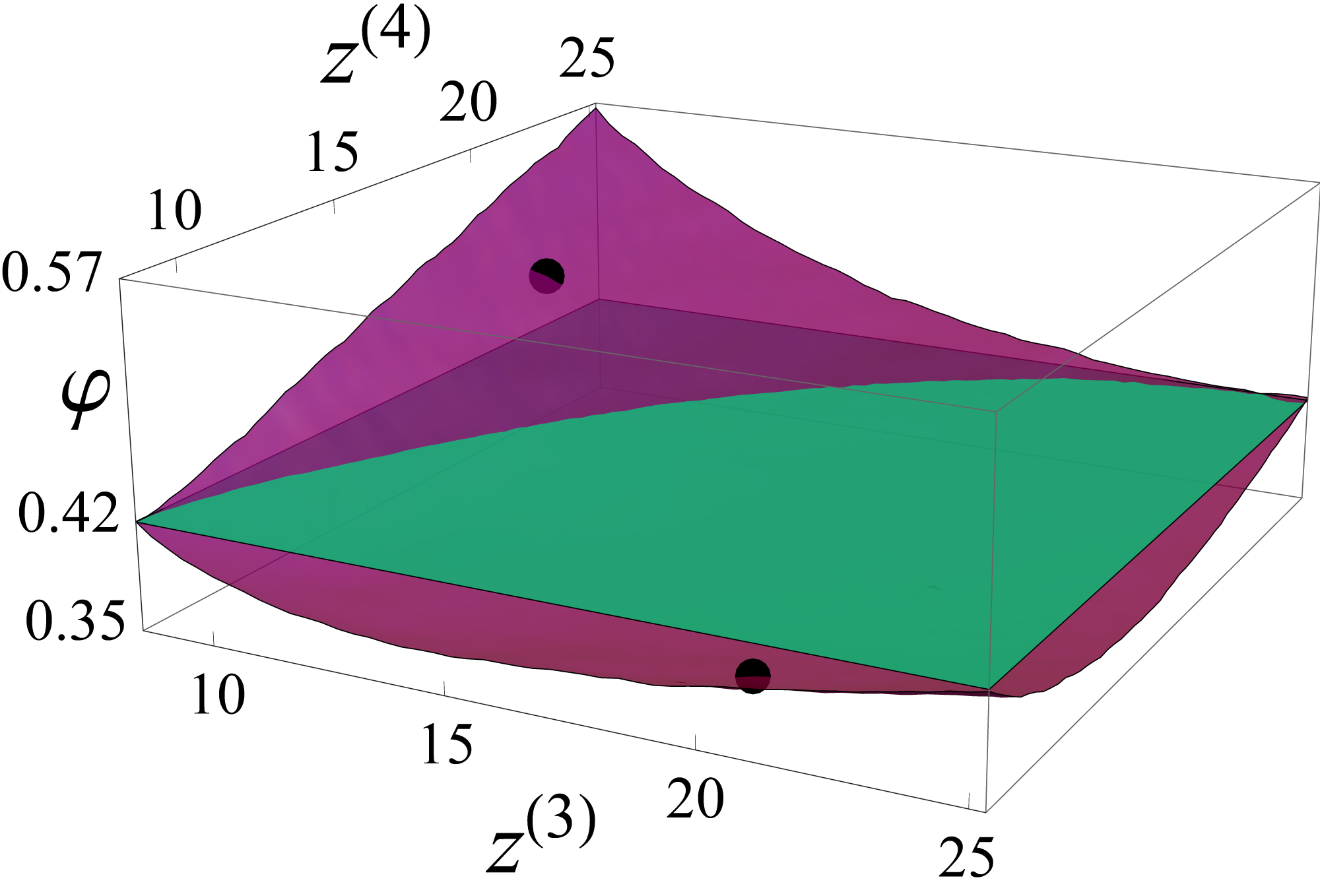}} \quad \sidesubfloat[]{\includegraphics[width=0.45\textwidth]{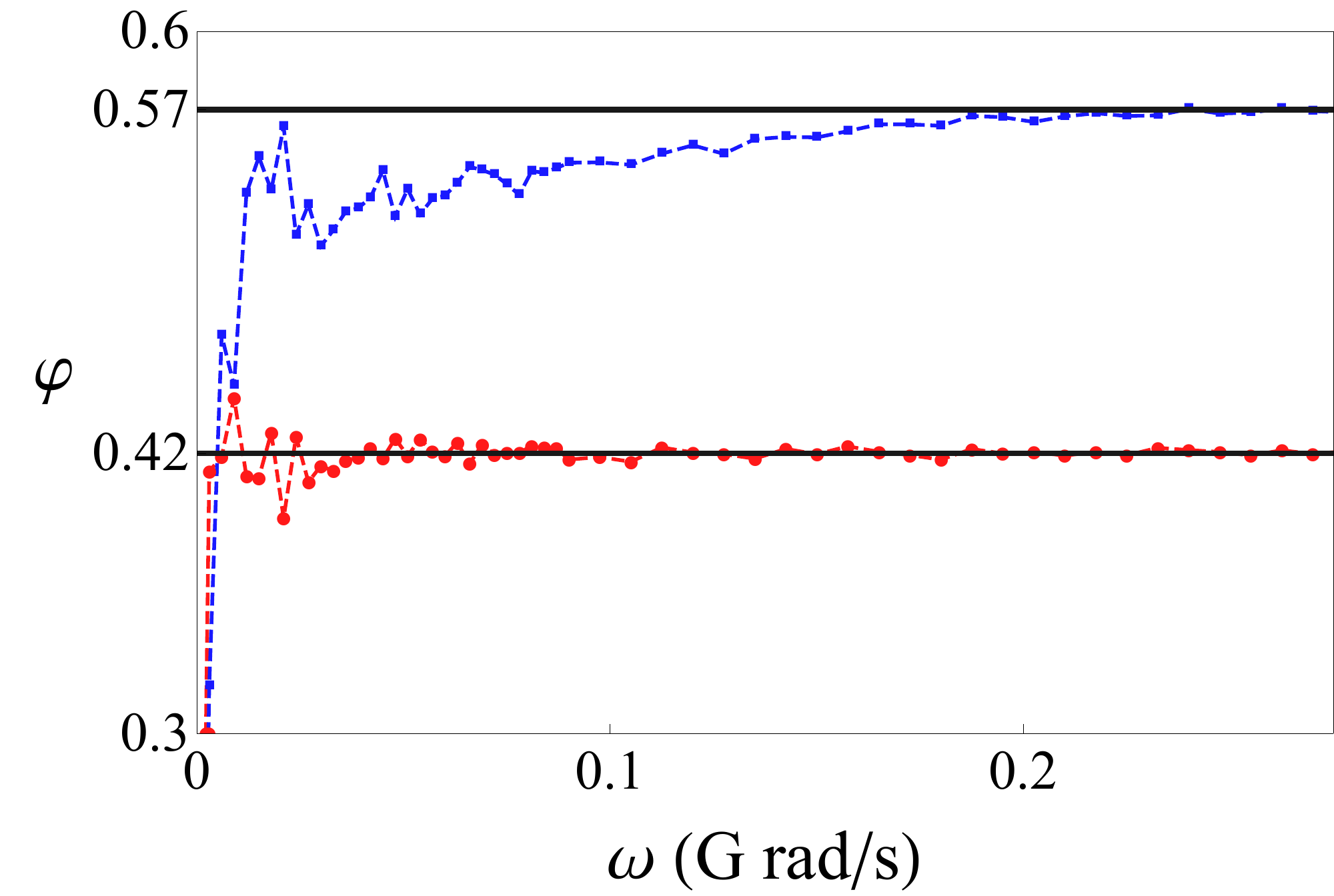}}\caption{(a) Gap density as function of layers 3 and 4 impedance. Layers 1
and 2 impedance is as in \eqref{eq:laminate 1}$_{1,2}$. The units
of the $\protect\nimp 3$ and $\protect\nimp 4$ axes are $\mathrm{M}\,\mathrm{kg\,m^{-2}\,s^{-1}}$
(b) The $\varphi_{\omega_{0}}$ sequences of 2-layer laminate (red
marks) with phases \eqref{eq:laminate 1}$_{1,2}$ and microstructure
\eqref{eq:thickness laminate 2 layers}, and a 4-layer laminate (blue
marks) of the same phases, with  microstructure \eqref{eq:thickness laminate 4 layers}.
Corresponding values of the gap density calculated over $\protect\ntorus 2$
and $\protect\ntorus 4$, respectively, are given by black lines.}

{\small{}{}\label{GD of z3z4}} 
\end{figure}

\emph{Laminates with 4 layers and more.} Fig. \ref{GD of z3z4}(a)
illustrates the gap density as function of layers 3 and 4 impedance,
when layers 1 and 2 impedance is as in \eqref{eq:laminate 1}. As
reference, the gap density of 2-layer laminates with the same $\gnm 12$
is represented by the green plane $\gd=0.42$. A comparison with laminates
whose unit cell comprises 3 layers shows two significant differences.
(i) Here, the set $\gset$ depends on the ordering of the phases,
while for 3-layer laminates it does not. Specifically, $\gnm{1,2}{3,4}$
quantifies impedance mismatch of layers 1 and 3 versus layers 2 and
4. It follows that interchanging layers, except layers 1 and 3 or
2 and 4, changes $\gnm{1,2}{3,4}$ and, in turn, the gap density.
Therefore, the bullet marked exemplary points $\left(10.07,20.22\right)$
and $\left(20.22,10.07\right)$ in Fig. \ref{GD of z3z4} have different
gap density. (ii) While the gap density of 2-layer laminates cannot
be improved by introducing a third layer if $\nimp 1<\nimp 3<\nimp 2$,
it can be improved by introducing a fourth layer, even when $\nimp 1<\left(\nimp 3,\nimp 4\right)<\nimp 2$.
Interestingly, the optimal composition is when layers 3 and 4 are
made of the phases that comprise layers 1 and 2, respectively, as
it yields the greatest $\gnm{1,2}{3,4}$; of course, the thickness
of layers 3 and 4 must be different than layers 1 and 2, and satisfy
Eq. \eqref{eq:irrationally independent}. This is demonstrated in
Fig. \ref{GD of z3z4}(b) by evaluating two $\varphi_{\omega_{0}}$
sequences, associated with two laminates made of phases \eqref{eq:laminate 1}$_{1,2}$,
where one laminate is composed of 2 layers (red marks) with \begin{equation} 
\begin{array}{llll}
\nthc 1=3\, \mathrm{mm}, & \nthc 2=2\, \mathrm{mm}, & & \\
\end{array}
\label{eq:thickness laminate 2 layers}
\end{equation}and the other is composed of 4 layers (blue marks), with 

\begin{equation} 
\begin{array}{llll}

\nthc 1=1.3\, \mathrm{mm}, & \nthc 2=1.3\, \mathrm{mm}, & \nthc 3=1.7\, \mathrm{mm}, & \nthc 4=1.7\, \mathrm{mm}. \\
\end{array}
\label{eq:thickness laminate 4 layers}
\end{equation}The corresponding calculations of the gap densities over $\ntorus 2$
and $\ntorus 4$, respectively, are given by the black lines. Indeed,
we observe that the sequences converge to the theoretical values,
and that the gap density of the 4-layer laminate is higher than the
density of the 2-layer laminate. 

\begin{figure}[t]
\centering\includegraphics[width=0.8\textwidth]{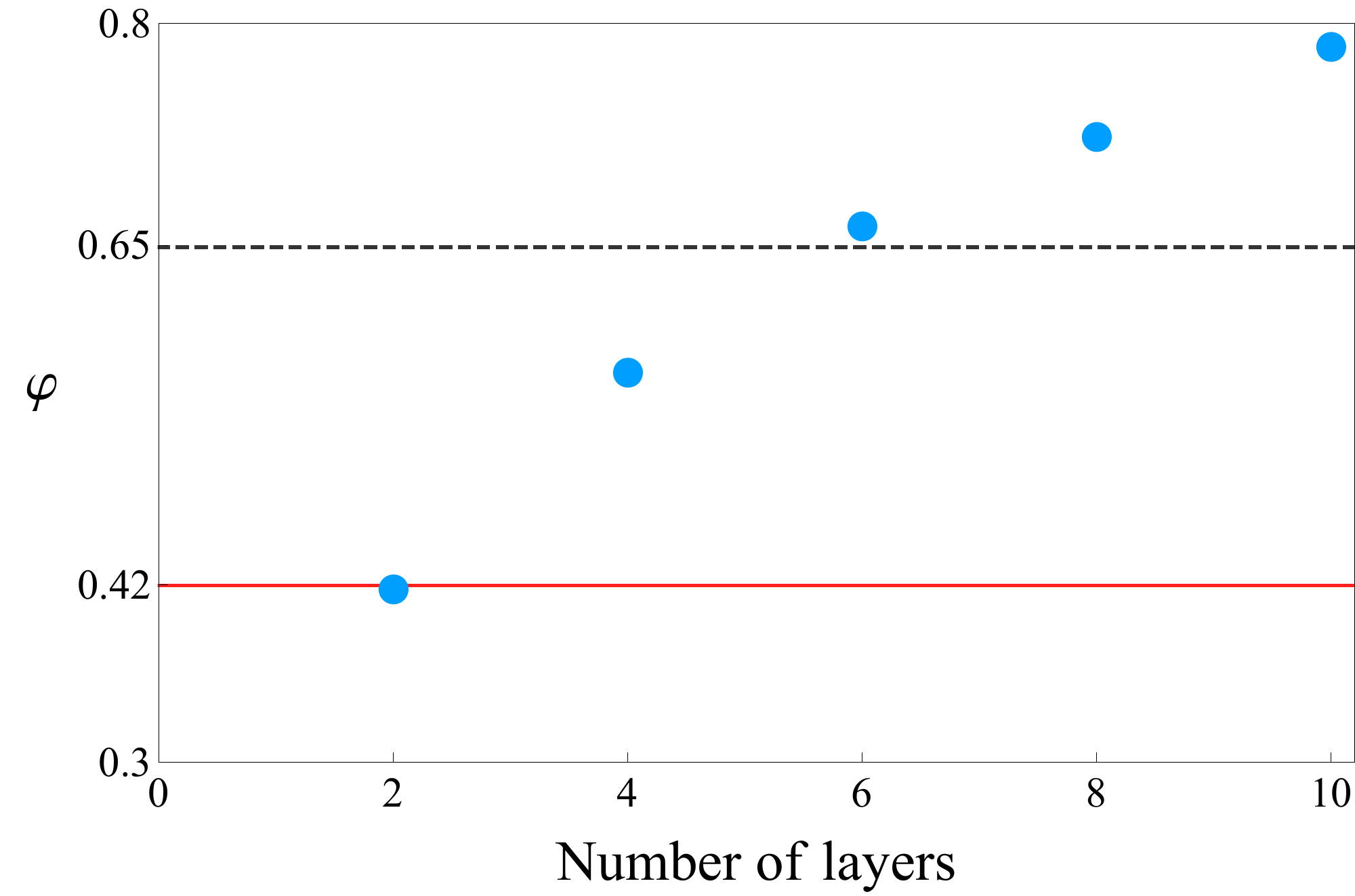}\caption{Gap density as function of the number of layers in a laminate made
of phases 1 and 2 as in \eqref{eq:laminate 1}, where the two-layer
gap density is shown also by the red solid line. The dashed line corresponds
to a two-layer laminate whose phase 1 (resp. 2) impedance is $3/2$
(resp. $2/3$) the impedance of phase 1 (resp. 2) in \eqref{eq:laminate 1}. }

{\small{}{}\label{GD 2 con vs multi}} 
\end{figure}
Assuming the laminate is subjected to a constraint on the impedance
range of available phases, our analysis suggests that the optimal
scheme to increase the gap density is by adding pairs of layers made
of the same two phases that constitute the 2-layer laminate, and varying
their thickness. We illustrate this in Fig. \ref{GD 2 con vs multi},
where the gap density is plotted against the number of layers in a
laminate made of phases 1 and 2 as in \eqref{eq:laminate 1}$_{1,2}$.
As reference, the 2-layer case is also shown in red line. For comparison,
the gap density of a 2-layer laminate whose phase 1 (resp. 2) impedance
is $2/3$ (resp. $3/2$) the impedance of phase 1 (resp. 2) in \eqref{eq:laminate 1}
is shown in dashed line. 

\subsection{\label{subsec:Applications-in-gap}Maximization of the $1^{\mathrm{st}}$
gap}

We apply next the general formulation in Sec. \ref{subsec:Gap-width-analysis}
to the address the following question. Consider two phases and the
2-layer microstructure that maximizes the $\fst$ gap at a prescribed
unit cell thickness $h$; can the $\fst$ gap be widened by introducing
a third layer made of a phase whose impedance is constrained between
the impedances of the other two phases?

To answer the question, we first analyze the 2-layer case. The $\fst$
gap corresponds to the segment originating at $\overrightarrow{b}=\overrightarrow{0}$.
Since the phase properties and $h$ are prescribed, the optimization
variable is only the thickness of one of the layers, say layer 1.
Over the torus, this translates to optimization over the possible
slopes $\nslop 1$. Using the expression for $\mathbb{D}^{2}$ and
its properties, we find that 
\begin{equation}
\max_{\nslop 1}\dz=\dz\left(\nslop 1=1\right)=2\sqrt{2}\arctan\frac{\sqrt{\gamma+\sqrt{\gamma^{^{2}}-1}}-\sqrt{\gamma-\sqrt{\gamma^{2}-1}}}{2},\label{eq:max dz 2 layer}
\end{equation}
where $\gamma=\gnm 12$. The second part of the bound also admits
an analytic solution, namely, 
\begin{equation}
\max_{0<h^{\left(1\right)}<h}\left(\frac{h^{\left(1\right)^{2}}}{c^{\left(1\right)^{2}}}+\frac{\left(h-h^{\left(1\right)}\right)^{2}}{c^{\left(2\right)^{2}}}\right)^{-\frac{1}{2}}=\frac{\sqrt{c^{\left(1\right)^{2}}+c^{\left(2\right)^{2}}}}{\thc}\label{eq:second max}
\end{equation}
at $h^{\left(1\right)}=\frac{c^{\left(1\right)^{2}}}{c^{\left(1\right)^{2}}+c^{\left(2\right)^{2}}}h$,
or in terms of the slope, at $\nslop 1=\frac{c^{\left(2\right)}}{c^{\left(1\right)}}$,
implying that the bound is sharp when the layers have the same phase
velocity.\floatsetup[figure]{style=plain,subcapbesideposition=top}
\begin{figure}[t]
\centering\sidesubfloat[]{\includegraphics[width=0.4\textwidth]{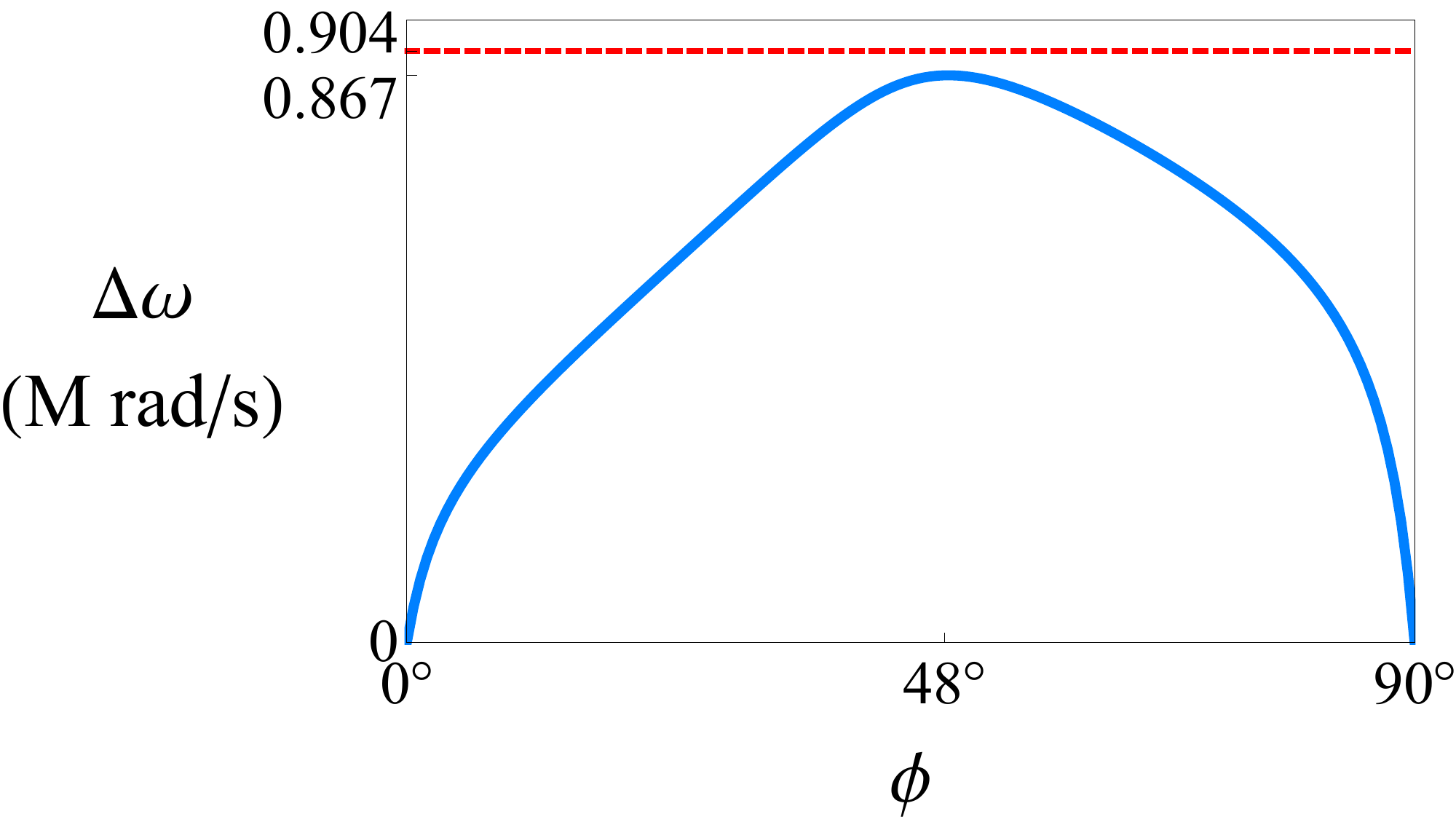}} \quad \sidesubfloat[]{\includegraphics[width=0.4\textwidth]{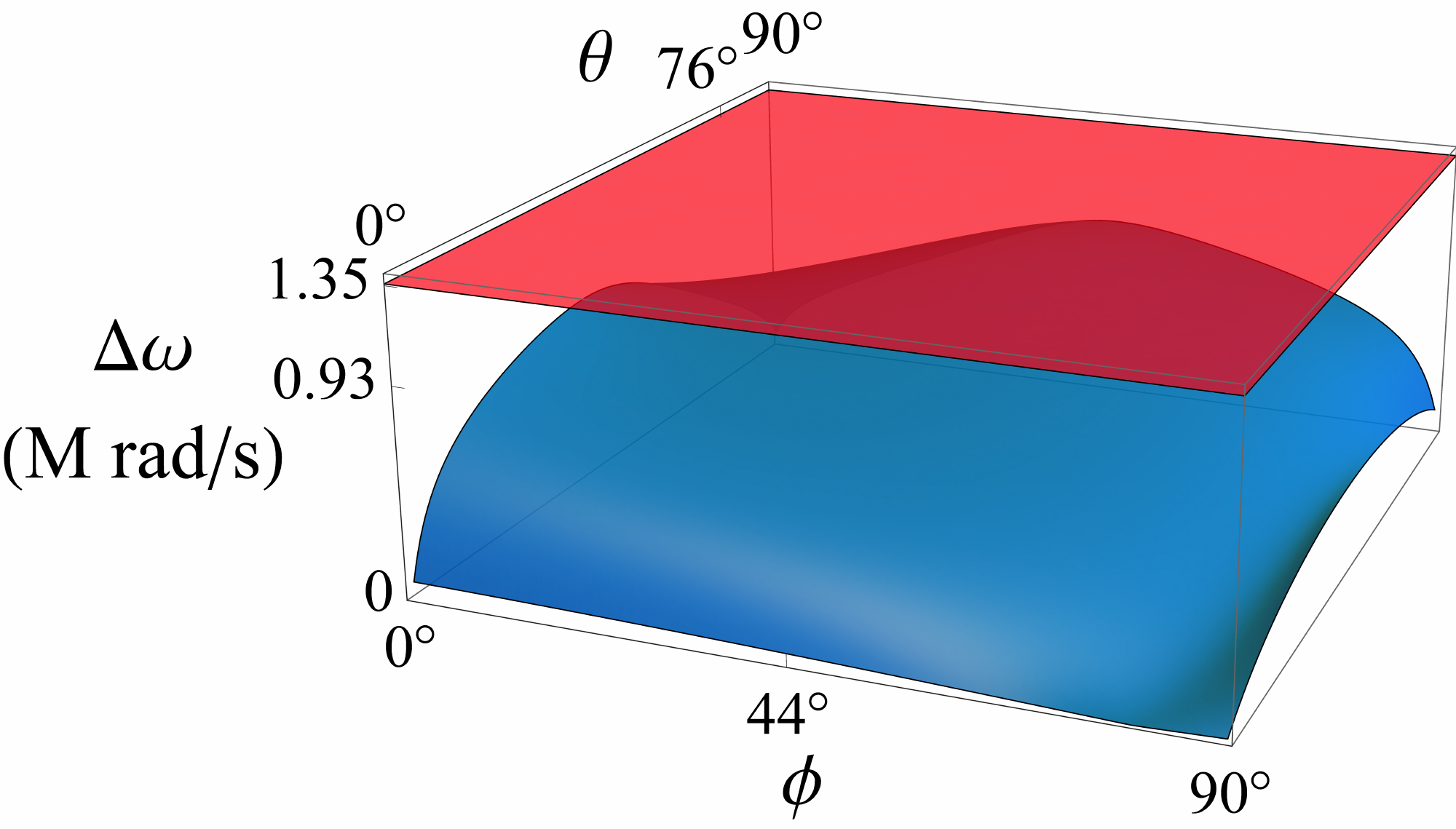}}\caption{(a) $\protect\fst$ gap width as function of the $\theta$ (continuous
blue curve), when phase 1 is \eqref{eq:constit laminate 4 and 5},
phase 2 is \eqref{eq:laminate 1}$_{2}$, and $h=11.5\,\mathrm{mm}$.
Our bound is depicted by the dashed red line. (b) $\protect\fst$
gap width (blue surface) as function of the possible microstructures
of a 3-layer in terms of $\phi$ and $\theta$, whose phase properties
are \eqref{eq:constit laminate 4 and 5}, \eqref{eq:laminate 1}$_{2}$,
and \eqref{eq: comp 3 for widest gap fig}. Our bound is depicted
by the red plane.}

{\small{}{}\label{Widest First Gap}} 
\end{figure}

We analyze next the 3-layer case. We start with the second part of
the bound. Given a third phase characterized by $c^{\left(3\right)}$,
the analytic solution reads
\begin{equation}
\max_{\overset{h^{\left(1\right)},h^{\left(2\right)}}{\mathrm{s.t.}\ 0<h^{\left(1\right)}+h^{\left(2\right)}<h}}\left(\frac{h^{\left(1\right)^{2}}}{c^{\left(1\right)^{2}}}+\frac{h^{\left(2\right)^{2}}}{c^{\left(2\right)^{2}}}+\frac{\left(h-h^{\left(1\right)}-h^{\left(2\right)}\right)^{2}}{c^{\left(3\right)^{2}}}\right)^{-\frac{1}{2}}=\frac{\sqrt{c^{\left(1\right)^{2}}+c^{\left(2\right)^{2}}+c^{\left(3\right)^{2}}}}{\thc}\label{eq:second max 3 layers}
\end{equation}
at $h^{\left(1\right)}=\frac{c^{\left(1\right)^{2}}}{c^{\left(1\right)^{2}}+c^{\left(2\right)^{2}}+c^{\left(3\right)^{2}}}h$
and $h^{\left(2\right)}=\frac{c^{\left(2\right)^{2}}}{c^{\left(1\right)^{2}}+c^{\left(2\right)^{2}}+c^{\left(3\right)^{2}}}h$,
or in terms of the slopes, at $\nslop 1=\frac{c^{\left(3\right)}}{c^{\left(1\right)}}$
and $\nslop 1=\frac{c^{\left(3\right)}}{c^{\left(2\right)}}$. If
the velocity $c^{\left(3\right)}$ is also a variable, the maximum
is achieved at the greatest admissible $c^{\left(3\right)}$. We have
employed a numerical analysis of $\mathbb{D}^{3}$ to find that $\max\dz$
is along the diagonal of the face associated with the greatest $\gnm nm$,
namely, 
\begin{equation}
\max\dz=2\sqrt{2}\max_{n,m}\left\{ \arctan\frac{\sqrt{\gnm nm+\sqrt{\gamma^{\left(n,m\right)^{2}}-1}}-\sqrt{\gnm nm-\sqrt{\gamma^{\left(n,m\right)^{2}}-1}}}{2}\right\} .\label{eq:max zeta 3}
\end{equation}

We evaluate next the above results in the following numerical example.
We set the properties of phase 1 to Eq. \eqref{eq:constit laminate 4 and 5},
phase 2 to Eq. \eqref{eq:laminate 1}$_{2}$, and the unit cell thickness
to $11.5\,\mathrm{mm}$. For these parameters, Eqs. \eqref{eq:do dz},
\eqref{eq:max dz 2 layer} and \eqref{eq:second max} provide the
bound $0.9\mathrm{04M}\,\frac{\mathrm{rad}}{\mathrm{s}}$. Fig. \ref{Widest First Gap}(a)
compares this bound (dashed red line) with the width as function of
the polar angle $\phi$, where $\tan\phi=\nslop 1$ (continuous blue
curve). The optimal value, namely, $0.867\mathrm{M}\,\frac{\mathrm{rad}}{\mathrm{s}}$
at $\phi=48^{\circ}$, found to be very close to the bound. Consider
a third phase whose properties are

\begin{equation}
\begin{array}{ll}
\nrho 3=3500\,\mathrm{kg/m^{3}} & \nmu 3=55\,\mathrm{GPa,}\end{array}\label{eq: comp 3 for widest gap fig}
\end{equation}
such that its impedance is between the impedances of phases 1 and
2. Eqs. \eqref{eq:do dz}, \eqref{eq:second max 3 layers} and \eqref{eq:max zeta 3}
yield the bound $1.35\,\frac{\mathrm{rad}}{\mathrm{s}}$ on the $1^{\mathrm{st}}$
gap width when a third layer made of the latter phase is added. Fig.
\ref{Widest First Gap}(b) compares this bound (red plane) with the
width (blue surface) as function of the possible microstructures of
the 3-layer in terms of $\phi$ and $\theta$, where $\tan\azi=\frac{\nslop 1}{\nslop 2}$
and $\tan\inclin=\sqrt{\left(1\slash\nslop 1\right)^{2}+\left(1\slash\nslop 2\right)^{2}}$.
 The maximal width observed is $0.93\mathrm{M}\,\frac{\mathrm{rad}}{\mathrm{s}}$;
indeed, as the change in the bound suggested, the $\fst$ gap can
be widened by introducing a third layer, even when its impedance is
subjected to constraint \eqref{eq:z1z3z2}. 

\section{\label{sec:Waves-superposed-on}Small-amplitude waves in finitely
deformed multiphase laminates}

Constitutively non-linear materials undergoing reversible large deformations
attract growing attention, since they can comply with varying functional
needs by virtue of their tunable geometrical and physical properties.
Clearly, such materials can be useful when it is desirable to obtain
tunable band diagrams, \emph{e.g.}, when the frequencies needed filtering
are changing \citep{bertoldi08prb,Wang2012,GETZ2017b,Barnwell2017parnell,gonella17jmps}.
The scheme for such tunability via finitely deformable multiphase
laminates is simple, and we start with its informal description.
The tuning mechanism is based upon an application of a finite quasi-static
deformation, say, by an axial force, which results in a change of
the thickness of each layer, and generally, its mass density. The
constitutive behavior of the phases is non-linear, thus their instantaneous
stiffness changes too. The propagation of small-amplitude waves depends
on the above quantities, which are rendered tunable by the finite
deformation; in turn, the band structure is rendered tunable too \citep{gg12}. 

\citet{Shmuel2016JMPS} employed the theory of incremental deformations
superposed on large deformations \citep{ogden97book,wavesbook07}
to show that the dispersion relation of finitely deformed laminates
made of two alternating layers is of the same functional form of Eq.
\eqref{eq:dispersion-N} for $N=2$. \citet{Shmuel2016JMPS} further
showed that the torus representation is applicable in this case too,
and employed the compactness and universality of the torus to characterize
the tunability. We concisely describe next the extension to finitely
deformed multiphase laminates.

Consider again the laminate in Sec. \ref{sec:Wave-propagation}, only
now the unit cell comprises $N$ \emph{non-linear} hyperelastic phases,
whose stress is derived from functions $\n{\aef}\left(\dg\right)$
of the finite deformation gradient $\dg$ via
\begin{equation}
\n{\cauchy}=\frac{1}{\det\dg}\frac{\partial\n{\Psi}}{\partial\mathbf{F}}\dg^{\mathrm{T}}.\label{eq:stress hyperelasticity}
\end{equation}
In addition to the usual physical requirements on $\left\{ \n{\aef}\right\} $,
we require symmetry with respect to the lamination direction $\nh=\mathbf{e}_{3}$,
but otherwise $\n{\aef}$ are of general form. The laminate is quasi-statically
deformed by applying an axial force of density $\fd$ per undeformed
unit area. To determine the resultant deformation, the equations of
continuity, far field conditions and equilibrium are to be satisfied.
To this end, we postulate a piecewise homogeneous deformation, such
that the deformation gradient matrix in layer $n$ is 

\begin{equation}
\n{\mathsf{F}}=\mathrm{diag}\left[\n{\lambda_{1}},\phase{\lambda_{2}},\phase{\lambda_{3}}\right],\label{eq:defromation gradient p f,m}
\end{equation}
namely, unit cubes comprising layer $n$ are deformed to $\n{\lambda_{1}}\times\phase{\lambda_{2}}\times\phase{\lambda_{3}}$
cuboids. The outstanding quasi-static problem to determine $\left\{ \n{\lambda_{1}},\phase{\lambda_{2}},\phase{\lambda_{3}}\right\} $
is  addressed as follows. Firstly, the in-plane stretches in the different
layers must match for the displacement field to be continuous across
the interfaces, hence
\begin{equation}
\fb{\lambda_{2}}=\mt{\lambda_{2}}=...=\N{\lambda_{2}}\eqqcolon\lambda_{2},\fb{\lambda_{3}}=\mt{\lambda_{3}}=...=\N{\lambda_{3}}\eqqcolon\lambda_{3}.\label{eq:in plane stretches}
\end{equation}
The problem symmetry in the plane implies that 
\begin{equation}
\lambda_{2}=\lambda_{3}\eqqcolon\lip,\label{eq:inplane stretch}
\end{equation}
and equilibrium determines the remaining relations. Specifically,
it leads to a constant axial stress throughout the laminate
\begin{equation}
\scomp{11}1=\scomp{11}2=...=\scomp{11}N=\frac{\fd}{\lip^{2}}.\label{eq:axial stress}
\end{equation}
Since the only force applied is axial, the remaining equation that
completes the set of $N+1$ equations for the stretches $\left\{ \phase{\lambda_{1}}\right\} $
and $\lip$ is
\begin{equation}
\sum_{n=1}^{N}\phase{\lambda_{1}}H^{\left(n\right)}\scomp{22}n=0,\label{eq:sigma bar}
\end{equation}
where $H^{\left(n\right)}$ is the thickness of layer $n$ before
the deformation. Note that for incompressible phases, the stretches
are constrained via $\lip^{2}\phase{\lambda_{1}}=1$, and a hydrostatic
Lagrange multiplier is added to the stress. This completes the calculation
of the laminate state in the deformed configuration.

To analyze small-amplitude wave propagation in the deformed laminate,
a linearization is carried out about the deformed configuration. In
each layer, we obtain a wave equation for the displacements $\phase{\mathbf{\mathbf{u}}}\left(\mathbf{x},t\right)$
from the \emph{deformed} state, namely, 
\begin{equation}
\divg\left(\phase{\elaspush}\grad\phase{\mathbf{u}}\right)=\phase{\rho}\phase{\ddot{\mathbf{u}}},\label{eq:inc equations-1}
\end{equation}
where the instantaneous elasticities are 
\begin{equation}
\phase{\elaspushcomp{ijkl}}=\frac{1}{\det\phase{\dg}}\phase{\dgcomp{j\alpha}}\derivative{^{2}\phase{\aef}}{\dgcomp{i\alpha}\partial\dgcomp{k\beta}}\phase{\dgcomp{l\beta}}.\label{eq:inc stress-1}
\end{equation}
Note that the current mass densities $\left\{ \n{\rho}\right\} $
depend on the finite deformation via $\n{\rho}=\n{\rho}_{0}/\det\n{\dg}$,
where $\n{\rho}_{0}$ is the initial mass density of phase $n$. It
follows that Eq. \eqref{eq:inc equations-1} admits solutions in the
form of Eq. \eqref{eq:uform-1}, and the resultant continuity and
Bloch-Floquet conditions reproduce Eq. \eqref{eq:dispersion-N}, only
now $\left\{ \n h,\n c,\n{\rho}\right\} $ \emph{are functions of}
$\fd$. Since $\eta$ maintains its functional form, the arguments
in Sec. \eqref{sec:A-universal-spectra} hold, hence the band structure
of small-amplitude waves in finitely deformed multiphase laminates
is also a linear flow on $\ntorus N$. The universality of the torus
representation offers a convenient platform to analyze how the finite
deformation affects the band structure, and replaces the need to carry
out numerous evaluations of band structures at different deformations.
This is demonstrated next, by way of examples. 

To avoid cumbersome calculations, we focus on anti-plane shear waves,
for which Eq. \eqref{eq:inc equations-1} simplifies to 
\begin{equation}
\phase{\elaspushcomp{2121}}\phase{u_{2,11}}=\phase{\rho}\phase{\ddot{u}_{2}}.\label{eq:anti-plane shear}
\end{equation}
The resultant coordinates on the torus are 
\begin{equation}
\n{\torusvar}=\frac{\omega\n h}{\n c}=\frac{\omega\n{\lambda}_{1}\n H}{\sqrt{\elaspushcomph{2121}^{\left(n\right)}/\phase{\rho_{0}}}},\ \left(\elaspushcomph{2121}^{\left(n\right)}=\det\phase{\dg}\phase{\elaspushcomp{2121}}\right)\label{eq:coordinates small on large}
\end{equation}
with the slopes 
\begin{equation}
\nslop n=\frac{\torusvar^{\left(N\right)}}{\torusvar^{\left(n\right)}}=\frac{h^{\left(N\right)}}{c^{\left(N\right)}}\frac{\n c}{\n h}=\frac{\N{\lambda}_{1}\N H}{\n{\lambda}_{1}\n H}\sqrt{\frac{\elaspushcomph{2121}^{\left(n\right)}\N{\rho_{0}}}{\elaspushcomph{2121}^{\left(N\right)}\phase{\rho_{0}}}},\label{eq:slopes small on large}
\end{equation}
and the impedance contrasts that determine $\left\{ \gamma\right\} $
are 
\begin{equation}
\frac{\zn m}{\zn n}=\frac{\rho^{\left(m\right)}c^{\left(m\right)}}{\rho^{\left(n\right)}c^{\left(n\right)}}=\frac{\lambda_{1}^{\left(n\right)}}{\lambda_{1}^{\left(m\right)}}\sqrt{\frac{\elaspushcomph{2121}^{\left(m\right)}\rho_{0}^{\left(m\right)}}{\elaspushcomph{2121}^{\left(n\right)}\n{\rho_{0}}}}.\label{eq:impedance constrasts small on large}
\end{equation}
To continue, the form of $\left\{ \phase{\aef}\right\} $ is needed.
Assuming incompressible Gent phases \citep{gent96rc&t}, the finite
deformation is homogeneous such that $\n{\lambda}_{1}=\lip^{-2}$,
and the instantaneous shear moduli are 
\begin{equation}
\elaspushcomph{2121}^{\left(n\right)}=\frac{\lip^{-4}\phase{\mu}}{1-\frac{\lip^{-4}+2\lip^{2}-3}{\p{\jm}}};\label{eq:shear moduli}
\end{equation}
here, $\phase{\mu}$ is the shear modulus of phase $n$ in the limit
of small strains, and $\n{\jm}\in(0,\infty]$ is a \emph{locking}
constant that models the stiffening of the material at finite strains.
We now can demonstrate how the tunability affects the flow and $\gap N$
in $\ntorus N$, depending on the relations between the constants
$\left\{ \n{\mu},\n{\jm}\right\} $, through the following two representative
cases. 
\begin{itemize}
\item Case 1: $\sup{\jm}1=\sup{\jm}2=...=\sup{\jm}N\eqqcolon\jm$. Under
this condition, the slopes $\left\{ \nslop n\right\} $ are independent
of the deformation, regardless of the relations between the shear
moduli and mass densities. The impedance contrasts do not change either,
 hence $\gap N$ and $\gd$ are invariant of the deformation. A change
is identified in the flow rate $\frac{\mathrm{d}\overrightarrow{\torusvar}}{\mathrm{d}\omega}$,
which now multiplies the factor $\sqrt{1-\frac{\lip^{-4}+2\lip^{2}-3}{\jm}}$,
and therefore the frequencies of each gap are divided by that factor.
Since this factor is smaller than $1$ and approaches $0$ as the
deformation becomes larger, the gaps are widened and shifted towards
higher frequencies. Note that the limit $\jm\rightarrow\infty$ corresponds
to \emph{neo-Hookean} phases\footnote{Note that the shear impedance contrasts between neo-Hookean phases
do not change even when the phases are compressible. In this case
we may have $\n{\lambda}_{1}\neq\lambda_{1}^{\left(m\right)}$ for
$n\neq m$ such that the the contrast between the velocities does
change, however this change is cancelled out in the calculation of
$\left\{ \gamma\right\} $ by the change of the density. This comment
corrects our error in \citet{Shmuel2016JMPS}, where we overlooked
this cancellation. }, whose analysis follows as a particular case, in which the flow rate
does not change either. 
\item Case 2: $\sup{\jm}1\neq\sup{\jm}2\neq...\neq\sup{\jm}N$. Contrarily
to the previous case, the layers stiffen at different rates. In turn,
the slopes and impedance contrasts change in a way that depends on
the relation between the different phase constants. There are numerous
classes of relations between the constants; exploring them is outside
our scope, however we provide an example for which the insights gained
in Sec. \ref{sec:Analysis-on-the} characterize certain aspects of
the tunability. Our example considers 3-layer laminates, whose phase
constants satisfy 
\begin{equation}
\sup{\mu}1=\sup{\mu}3<\sup{\mu}2,\ \jm^{\left(3\right)}<\jm^{\left(1\right)}=\jm^{\left(2\right)},\label{eq:impedance 3 layer case}
\end{equation}
where, for simplicity, their mass density is equal. In the undeformed
state, the band structure is identical to the band structure of a
2-layer laminate, encapsulated in $\ntorus 2$ with $\gamma=\gnm 12$.
Since $\jm^{\left(3\right)}<\jm^{\left(1\right)}$, the instantaneous
stiffness and (consequently) impedance of layer 3 exceeds that of
layer 1 upon deformation; up to a critical deformation, the resultant
instantaneous impedances satisfy relation \eqref{eq:z1z3z2}, with
$\max_{n,m}\gnm nm$ fixed, and equal to $\gnm 12$. The band structure
is described on $\ntorus 3$, and our torus analysis in Sec. \ref{subsec:3-layer-laminates}
shows that at fixed $\max_{n,m}\gnm nm$, the gap density in $\ntorus 3$
is always lower than in $\ntorus 2$ of the same $\gamma$. Thus,
we deduce that the effect deformation has on the band structure is,
on average, to narrow the gaps relatively to the undeformed state.
At the critical deformation in which the strains are large enough
to satisfy 
\begin{equation}
\frac{\lip^{-4}+2\lip^{2}-3-\jm^{\left(3\right)}}{\lip^{-4}+2\lip^{2}-3-\jm^{\left(2\right)}}\frac{\jm^{\left(2\right)}}{\jm^{\left(3\right)}}=\frac{\sup{\mu}3}{\sup{\mu}2},\label{eq:impedance ratio z3z2}
\end{equation}
we have that $\gnm 13=\gnm 12$, hence the band structure is of a
2-layer laminate whose gap density equals the gap density of the undeformed
laminate. Beyond this loading point, the instantaneous maximal impedance
mismatch becomes greater than at lower strains. Our torus analysis
in Sec. \ref{subsec:3-layer-laminates} indicates that, in turn, the
resultant gap density is greater than the gap density at the undeformed
state.
\begin{figure}[t]
\centering\includegraphics[width=0.8\textwidth]{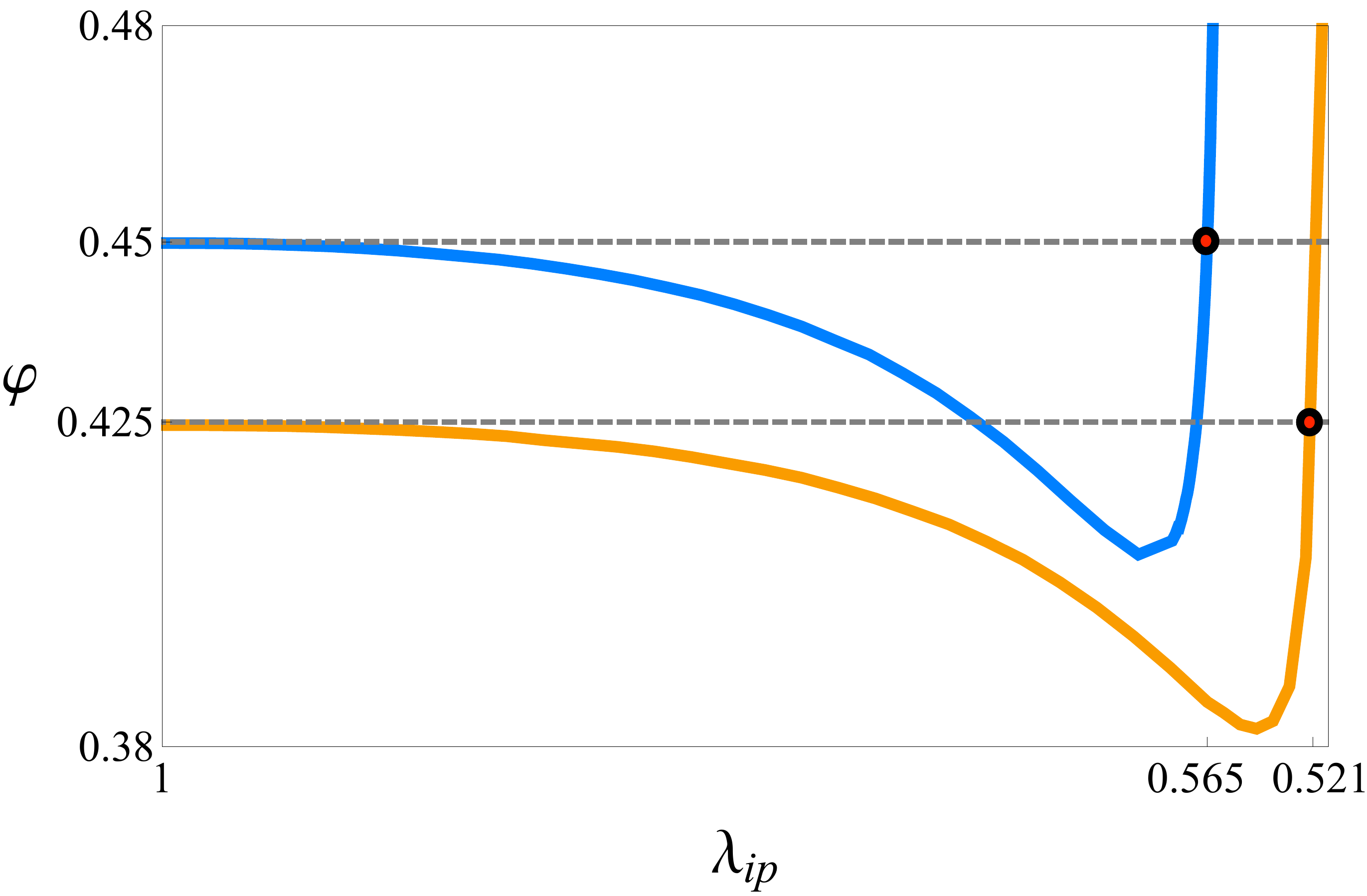}\caption{Gap density as function of the pre-deformation parameter $\protect\lip$,
for the sets of materials . \eqref{eq:laminate small on large set1}
and \eqref{eq:laminate small on large set2}, depicted in blue and
orange curves, respectively. }

{\small{}{}\label{Small On Large}} 
\end{figure}
\end{itemize}
We conclude this Sec. with a numerical calculation demonstrating the
latter result, considering 3-layer laminates whose properties are
characteristic properties of elastomers \citealp[(see, \emph{e.g.,}][ and the references therein)]{Getz2016gsm}.
Specifically, the two laminates considered are given by \begin{equation}
\begin{array}{lll}
\nrho 1=\nrho 2=\nrho 3=1000\,\mathrm{kg/m^{3}},
\\

\nmu 1=\nmu 3=500,\ \nmu 2=5500\, \mathrm{kPa}, 
\\
\njm 1=\njm 2=35,\ \njm 3=8,
\end{array}
\label{eq:laminate small on large set1}
\end{equation}and \begin{equation}
\begin{array}{lll}
\nrho 1=\nrho 2=\nrho 3=1000\,\mathrm{kg/m^{3}},
\\

\nmu 1=\nmu 3=650,\ \nmu 2=6100\, \mathrm{kPa}, 
\\
\njm 1=\njm 2=40,\ \njm 3=12.
\end{array}
\label{eq:laminate small on large set2}
\end{equation}The gap density of laminates \eqref{eq:laminate small on large set1}
and \eqref{eq:laminate small on large set2} as function of the pre-deformation
parameter $\lip$ is depicted Fig. \ref{Small On Large} by the blue
and orange curves, respectively. Indeed, as our analysis implied,
when the layers satisfy Eq. \eqref{eq:impedance 3 layer case}, pre-deformations
reduce the gap density in comparison with the gap density of the undeformed
laminates. At the critical deformation $\lip=0.565$, the instantaneous
moduli of layers 2 and 3 in laminate \eqref{eq:laminate small on large set1}
are equal, namely, $\elaspushcomph{2121}^{\left(2\right)}=\elaspushcomph{2121}^{\left(3\right)}=68.3\,\mathrm{MPa}$.
Accordingly, at this deformation the gap density equals the gap density
of the undeformed laminate, as indicated by the red dot in the plot.
The maximal impedance mismatch beyond this point exceeds its counterpart
at $\lip=1$, hence the gap density is greater than in the reference
configuration. Laminate \eqref{eq:laminate small on large set2} undergoes
a similar process, with the critical deformation $\lambda_{ip}=0.521$,
at which $\elaspushcomph{2121}^{\left(2\right)}=\elaspushcomph{2121}^{\left(3\right)}=114.1\,\mathrm{MPa}$.

\section{\label{sec:Concluding-remarks}Summary }

We have shown that the infinite band diagrams of waves at normal incident
angle to \emph{N}-phase laminates are encapsulated in a universal
compact manifold, namely, $\ntorus N$, independently of the specific
properties of the layers. The frequency parametrizes a linear flow
over $\ntorus N$, whose intersection with the submanifold $\gap N$
is identified with gaps in the band diagrams. Using ergodicity arguments,
we proved that the gap density of \emph{N}-phase laminates is a well-defined
quantity, universal for classes of laminates, and equals $\frac{\mathrm{vol}\gap N}{\mathrm{vol}\ntorus N}.$
This result was supplemented with numerical examples, demonstrating
that the calculation of the gap density in the original band structures,
truncated at increasing frequencies for different multiphase laminates,
converges to the relative volume of $\gap N$.

We have employed our theory in a numerical study on the relation between
the gap density, number of layers, and their impedance. Interestingly,
we found that the gap density of two-layer laminates is increased
by adding to the unit cell pairs of layers made of the same phases
at different thicknesses. We have also showed that the encapsulation
of the band diagrams in $\ntorus N$ is useful for formulating optimization
problems on the gaps width, and developed a simple bound. By way of
example, we addressed the question: can we enlarge the $\fst$ gap
of the optimal two-layer laminate by adding phases to the unit cell,
even when the impedance of the added phases is constrained between
the impedance of the original laminate phases? To this end, we have
analytically calculated the bound for two-layer laminates, finding
it is sharp. We have semi-analytically calculated the bound for the
constrained 3-layer laminate, and found it is higher than the bound
in the two-layer case, suggesting that the answer is yes. We have
further evaluated the gap width as function of the microstructure
for exemplary phases, using its expression over the torus in the two
cases. Indeed, this evaluation confirmed that the optimal microstructure
in the 3-layer case yields a wider $1^{\mathrm{st}}$ gap than the
optimal microstructure in the two-layer case. 

Finally, we have analyzed laminates comprising an arbitrary number
of non-linear finitely deformed phases. We have showed that if such
laminates undergo finite piecewise-constant deformations, the dispersion
relation of superposed small-amplitude waves is similar to the dispersion
relation of linear laminates, where the phase moduli and thickness
become functions of the finite deformation. Hence, our theory and
analysis of linear laminates extend to non-linear laminates under
the foregoing settings. In addition, we have demonstrated how our
framework facilitates the characterization of the band diagram tunability
by static finite deformations.

We conclude this paper noting that our theory applies for additional
multiphase systems, other than elastic laminates, whose dispersion
relation is functionally similar to Eq. \eqref{eq:dispersion-N},
\emph{e.g.}, certain electroelastic systems \citep{Qian04_eq,gg12}
and stratified photonic crystals \citep{gb16}, owing to the similarly
with the problem of electromagnetic waves \citep{Lekner94,adams08}. 

\section*{Acknowledgments}

An anonymous reviewer of a previous paper \citep{Shmuel2016JMPS}
is gratefully acknowledged on a comment made that intrigued our interest
in the present topic. We are also grateful to the support of the Israel
Science Foundation, funded by the Israel Academy of Sciences and Humanities
(Grant no. 1912/15), and the United States-Israel Binational Science
Foundation (Grant no. 2014358). 

\bibliographystyle{plainnat}

\end{document}